\documentclass[ALICE,manyauthors]{cernphprep}
\usepackage[comma,square,numbers,sort&compress]{natbib}
\usepackage{hyperref}
\usepackage{lineno}
\usepackage{xspace}
\usepackage{xcolor}
\usepackage{comment}
\usepackage{rotating}
\usepackage[utf8]{inputenc}
\usepackage{amsmath}
\usepackage[usestackEOL]{stackengine}
\usepackage[T1]{fontenc}
\usepackage{orcidlink}

\begin{document}
\newcommand{ \be }{\begin{linenomath*}\begin{eqnarray}}
\newcommand{ \ee }{\end{eqnarray}\end{linenomath*}}
\newcommand{ \la }{\langle}
\newcommand{ \lla }{\left\langle}
\newcommand{ \ra }{\rangle}
\newcommand{ \rra }{\right\rangle}
\def \mean#1 {{\la #1 \ra}}
\newcommand{ \dnavg}{{\ensuremath{\langle
    \mathrm{d}N_\mathrm{ch}/\mathrm{d}\eta \rangle}\xspace}_{|\eta|<0.5}}

\newcommand{ \pp}{pp}
\newcommand{ \pA}{p--A}
\newcommand{ \pPb}{p--Pb}
\newcommand{ \PbPb}{Pb--Pb}
\newcommand{ \AonA}{A--A}

\newcommand{ \pvec}{\vec{p}}
\newcommand{ \pvecOne}{\vec{p}_1}
\newcommand{ \pvecTwo}{\vec{p}_2}
\newcommand{ \pt }{\textit{p}_{\rm T}}
\newcommand{ \gevc }{\rm{GeV}/\textit{c}}

\newcommand{ \ptOne}{\textit{p}_{\rm T,1}}
\newcommand{ \ptTwo}{\textit{p}_{\rm T,2}}
\newcommand{ \pti}{\textit{p}_{\rm T,i}}
\newcommand{ \ptj}{\textit{p}_{\rm T,j}}
\newcommand{ \ptk}{\textit{p}_{\rm T,k}}
\newcommand{ \ptik}{\textit{p}_{\rm T,i,k}}
\newcommand{ \dpt }{\rm{d}\textit{p}_{\rm T}}
\newcommand{ \dptOne}{\rm{d}\textit{p}_{\rm T,1}}
\newcommand{ \dptTwo}{\rm{d}\textit{p}_{\rm T,2}}
\newcommand{ \dpti}{\rm{d}\textit{p}_{\rm T,i}}
\newcommand{ \dptj}{\rm{d}\textit{p}_{\rm T,j}}
\newcommand{ \Dpt }{\Delta \textit{p}_{\rm T}}
\newcommand{ \DptDpt }{\Delta \pt \Delta \pt }
\newcommand{ \DeltaPtPt}{\langle  \Delta \pt \Delta \pt \rangle}
\newcommand{ \DptOne}{\Delta \textit{p}_{\rm T,1}}
\newcommand{ \DptTwo}{\Delta \textit{p}_{\rm T,2}}

\newcommand{ \dphi}{\rm{d}\varphi}
\newcommand{ \phiOne}{\varphi_1}
\newcommand{ \phiTwo}{\varphi_2}
\newcommand{ \phii}{\varphi_i}
\newcommand{ \phij}{\varphi_j}
\newcommand{ \dphiOne}{\rm{d}\varphi_1}
\newcommand{ \dphiTwo}{\rm{d}\varphi_2}
\newcommand{ \Dphi }{\Delta \varphi}

\newcommand{ \etaOne }{\eta_1}
\newcommand{ \etaTwo }{\eta_2}
\newcommand{ \etai }{\eta_i}
\newcommand{ \etaj }{\eta_j}
\newcommand{ \deta }{\rm{d}\eta}
\newcommand{ \detaOne }{\rm{d}\eta_1}
\newcommand{ \detaTwo }{\rm{d}\eta_2}
\newcommand{ \Deta }{\Delta \eta}
\newcommand{ \etaPhi }{\left( \eta,\varphi \right)}
\newcommand{ \etaPhiI}{\left( \eta_i,\varphi_i\right)}
\newcommand{ \etaPhiOne}{\left( \etaOne,\phiOne\right)}
\newcommand{ \etaPhiTwo}{\left( \etaTwo,\phiTwo\right)}
\newcommand{ \etaPhiPt }{\left( \eta,\varphi,p_t \right)}
\newcommand{ \etaPhiPtI}{\left( \eta_i,\varphi_i,\textit{p}_{\rm T,i} \right)}
\newcommand{ \etaPhiPtOne}{\left( \etaOne,\phiOne,\textit{p}_{\rm T,1} \right)}
\newcommand{ \etaPhiPtTwo}{\left( \etaTwo,\phiTwo,\textit{p}_{\rm T,2} \right)}
\newcommand{ \etaPhiEtaPhi }{\left(  \etaOne,\phiOne,\etaTwo,\phiTwo \right)}
\newcommand{ \etaPhiPtEtaPhiPt }{\left(  \etaOne,\phiOne, \textit{p}_{\rm T,1}
	\etaTwo,\phiTwo, \textit{p}_{\rm T,2} \right) } 
\newcommand{ \DetaDphi }{\left(  \Delta \eta, \Delta \varphi \right)}
\newcommand{ \rhoOne}{\rho_1}
\newcommand{ \rhoTwo}{\rho_2}
\newcommand{\Rtwo}{R_2}
\newcommand{\Ptwo}{P_2}
\newcommand{\RtwoLS}{R_2^{\mathrm{LS}}}
\newcommand{\RtwoUS}{R_2^{\mathrm{US}}}
\newcommand{\RtwoCI}{R_2^{\mathrm{CI}}}
\newcommand{\RtwoCD}{R_2^{\mathrm{CD}}}
\newcommand{\PtwoLS}{P_2^{\mathrm{LS}}}
\newcommand{\PtwoUS}{P_2^{\mathrm{US}}}
\newcommand{\PtwoCI}{P_2^{\mathrm{CI}}}
\newcommand{\PtwoCD}{P_2^{\mathrm{CD}}}

\newcommand{\R}{\Rtwo}
\newcommand{\Rpt}{R^{(\Delta p_t)}_2}
\newcommand{\rhoPtPt}{\la \textit{p}_{\rm T} \textit{p}_{\rm T} \ra }
\newcommand{\rhoDptDpt}{\la \Delta \textit{p}_{\rm T}  \Delta \textit{p}_{\rm T} \ra }
\newcommand{\rpt}{\Rpt}

\newcommand{ \ds }{\displaystyle}
\newcommand{ \eps}{\varepsilon}
\newcommand{\empha}{et al.}
\newcommand{\dedx}{d$E$/d$x$}

\newcommand{\orange}[1]{\textcolor{orange}{#1}}
\newcommand{\blue}[1]{{\color{blue}{#1}}}
\newcommand{\green}[1]{{\color{dgreen}{#1}}}
\newcommand{\red}[1]{{\color{red}{#1}}}
\newcommand{\magenta}[1]{{\color{magenta}{#1}}}
\newcommand{\cyan}[1]{{\color{cyan}{#1}}}
\def \new {\blue}
\def \old {\orange}
\def \ask {\magenta}
\def \changed {\green}

\def \s {\sqrt{\textit{s}}}
\def \snn {\sqrt{\textit{s}_{_{\rm NN}}}}
\def \bp  {{\bf p}}
\def \mod {{\rm mod}}
\long\def\/*#1*/{}

\newcommand{\pipm}{\ensuremath{\mathrm{\pi}^{+}\mathrm{\pi}^{-}}\xspace}
\newcommand{\KP}{\ensuremath{\mathrm{K}^{\pm}\mathrm{\pi}^{\mp}}\xspace}
\newcommand{\KS}{(\ensuremath{\mathrm{K^{*0}}\rightarrow \mathrm{K}^{\pm}\mathrm{\pi}^{\pm}})\xspace}
\newcommand{\Ks}{\ensuremath{\mathrm{K} ^{*0}}\xspace}
\newcommand{\ptt}{\ensuremath{p_{\mathrm{T}}}\xspace}
\newcommand{\mtt}{\ensuremath{m_{\mathrm{T}}}\xspace}
\newcommand{\pb}{\mbox{Pb--Pb}\xspace}
\newcommand{\au}{\mbox{Au--Au}\xspace}
\newcommand{\ada}{\mbox{A--A}\xspace}

\newcommand{\rs}[1][8.0~TeV]{\mbox{\ensuremath{\sqrt{s}=} #1}\xspace}
\newcommand{\rso}{\mbox{\ensuremath{\sqrt{s}}}\xspace}
\newcommand{\rsnn}[1][2.76 TeV]{\mbox{\ensuremath{\sqrt{s_{\mathrm{NN}}}=} #1}\xspace}
\newcommand{\rsnno}{\mbox{\ensuremath{\sqrt{s_{\mathrm{NN}}}}}\xspace}

\newcommand{\gvc}{\mbox{\rm GeV$\kern-0.15em /\kern-0.12em c$}\xspace}
\newcommand{\gvcc}{\mbox{\rm GeV$\kern-0.15em /\kern-0.12em c^2$}\xspace}
\newcommand{\mvc}{\mbox{\rm MeV$\kern-0.15em /\kern-0.12em c$}\xspace}
\newcommand{\mvcc}{\mbox{\rm MeV$\kern-0.15em /\kern-0.12em c^2$}\xspace}

\newcommand{\pion}{\ensuremath{\uppi}\xspace}
\newcommand{\pix}{\ensuremath{\pion^{\pm}}\xspace}
\newcommand{\pim}{\ensuremath{\pion^{-}}\xspace}
\newcommand{\pip}{\ensuremath{\pion^{+}}\xspace}

\newcommand{\kx}{\ensuremath{\mathrm{K}^{\pm}}\xspace}
\newcommand{\km}{\ensuremath{\mathrm{K}^{-}}\xspace}
\newcommand{\kp}{\ensuremath{\mathrm{K}^{+}}\xspace}

\newcommand{\lab}{\ensuremath{(\Lambda+\overline{\Lambda})/2}\xspace}

\newcommand{\xim}{\ensuremath{\Xi^{-}}\xspace}
\newcommand{\xip}{\ensuremath{\overline{\Xi}^{+}}\xspace}
\newcommand{\xix}{\ensuremath{\xim+\xip}\xspace}

\newcommand{\omm}{\ensuremath{\Omega^{-}}\xspace}
\newcommand{\omp}{\ensuremath{\overline{\Omega}^{+}}\xspace}
\newcommand{\omx}{\ensuremath{\omm+\omp}\xspace}

\newcommand{\ks}{\ensuremath{\mathrm{K_{s}^{0}}}\xspace}
\newcommand{\ksb}{\ensuremath{\mathrm{\overline{K}^{0}}}\xspace}
\newcommand{\ksm}{\ensuremath{\mathrm{K^{*}(892)^{0}}}\xspace}
\newcommand{\ksbm}{\ensuremath{\mathrm{\overline{K}^{*}(892)^{0}}}\xspace}
\newcommand{\ph}{\ensuremath{\upphi}\xspace}
\newcommand{\phm}{\ensuremath{\ph(1020)}\xspace}

\newcommand{\dd}{\ensuremath{\mathrm{d}}\xspace}
\newcommand{\ddn}{\ensuremath{\dd^{2}N\kern-0.15em /\kern-0.12em\dd\ptt\dd y}\xspace}
\newcommand{\dddn}{\ensuremath{E\dd^{3}N\kern-0.15em /\kern-0.12em\dd p^{3}}\xspace}
\newcommand{\dndy}{\ensuremath{\dd N\kern-0.15em /\kern-0.12em\dd y}\xspace}
\newcommand{\npart}{\ensuremath{\langle N_{\mathrm{part}}\rangle}\xspace}
\newcommand{\ncoll}{\ensuremath{\langle N_{\mathrm{coll}}\rangle}\xspace}
\newcommand{\mult}{\ensuremath{\dd N_{\mathrm{ch}}\kern-0.06em /\kern-0.13em\dd\eta}\xspace}
\newcommand{\dncr}{\ensuremath{(\mult)^{1/3}}\xspace}
\newcommand{\mpt}{\ensuremath{\langle\ptt\rangle}\xspace}

\newcommand{\effr}{\ensuremath{\varepsilon_{\mathrm{rec}}}\xspace}
\newcommand{\effp}{\ensuremath{\varepsilon_{\mathrm{PID}}}\xspace}
\newcommand{\effpc}{\ensuremath{\varepsilon_{\mathrm{PC}}}\xspace}

\newcommand{\ncr}{\ensuremath{N_{\mathrm{cr,TPC}}}\xspace}
\newcommand{\rtpc}{\ensuremath{R_{\mathrm{TPC}}}\xspace}
\newcommand{\dcaxy}{\ensuremath{\mathrm{DCA}_{xy}}\xspace}
\newcommand{\dcaz}{\ensuremath{\mathrm{DCA}_{z}}\xspace}
\newcommand{\sktpc}{\ensuremath{\sigma_{\mathrm{K,TPC}}}\xspace}
\newcommand{\sktof}{\ensuremath{\sigma_{\mathrm{K,TOF}}}\xspace}
\newcommand{\spitpc}{\ensuremath{\sigma_{\mathrm{\pi,TPC}}}\xspace}
\newcommand{\spitof}{\ensuremath{\sigma_{\mathrm{\pi,TOF}}}\xspace}
\newcommand{\nsktpc}{\ensuremath{n\sktpc}\xspace}
\newcommand{\nsktof}{\ensuremath{n\sktof}\xspace}
\newcommand{\npiktpc}{\ensuremath{n\spitpc}\xspace}
\newcommand{\nspitof}{\ensuremath{n\spitof}\xspace}
\newcommand{\imin}{\ensuremath{I_{\mathrm{min}}}\xspace}
\newcommand{\imax}{\ensuremath{I_{\mathrm{max}}}\xspace}
\newcommand{\fnorm}{\ensuremath{f_{\mathrm{norm}}}\xspace}
\newcommand{\fvtx}{\ensuremath{f_{\mathrm{vtx}}}\xspace}

\newcommand{\sigc}{\ensuremath{\sigma_{\mathrm{central}}}\xspace}
\newcommand{\sigl}{\ensuremath{\sigma_{\mathrm{low}}}\xspace}
\newcommand{\sigh}{\ensuremath{\sigma_{\mathrm{high}}}\xspace}
\newcommand{\scc}{\ensuremath{\sigma_{\mathrm{cc}}}\xspace}
\newcommand{\dm}{\ensuremath{\langle\Delta M\rangle}\xspace}

\newcommand{\etal}{\textit{et al. }}


\begin{titlepage}
\PHyear{2024}       
\PHnumber{271}      
\PHdate{15 October}  
\title{Measurements of differential two-particle number and
  transverse momentum correlation functions in pp
  collisions at $\sqrt{\textit{s}}$ = 13 TeV}

\ShortTitle{$\Rtwo\DetaDphi$ and $\Ptwo\DetaDphi$ in pp @ 13 TeV}   

\Collaboration{ALICE Collaboration\thanks{See Appendix~\ref{app:collab} for the list of collaboration members}}
\ShortAuthor{ALICE Collaboration} 

\begin{abstract}
Differential two-particle normalized cumulants 
($R_2$) and transverse momentum correlations ($P_2$) are measured as a
function of the relative pseudorapidity and azimuthal angle difference
$( \Delta \eta, \Delta \varphi )$ of charged particle pairs in minimum bias pp collisions 
at $\sqrt{\textit{s}}$ = 13 TeV. The measurements use charged hadrons in the pseudorapidity region of $|\eta| < 0.8$ and the transverse momentum range \mbox{0.2 $< \textit{p}_{\mathrm T} < $ 2.0 $\mathrm{GeV}/\textit{c}$} in order to focus on soft multiparticle interactions and to complement  prior measurements of these correlation functions in p--Pb and Pb--Pb collisions. The correlation functions are reported for both unlike-sign and like-sign pairs and their charge-independent and charge-dependent combinations. Both the $R_2$ and $P_2$ measured in pp collisions exhibit  features qualitatively similar to those observed in p--Pb and Pb--Pb collisions. The $\Delta\eta$ and $\Delta\varphi$ root mean square widths of the near-side peak of the correlation functions are evaluated and compared with those observed  in p--Pb and Pb--Pb collisions and show smooth evolution with the multiplicity of charged particles produced in the collision. The comparison of the measured correlation functions with predictions from PYTHIA8 shows that this model qualitatively captures their basic structure and characteristics but feature important differences. In addition, the $R_2^{\rm CD}$ is used to determine the charge balance function of hadrons produced within the detector acceptance of the measurements. The integral of the balance function is found to be compatible with those reported by a previous measurement in  Pb--Pb collisions. 
\end{abstract}
\end{titlepage}

\setcounter{page}{2} 


\section{Introduction}
\label{sec:Introduction}
Understanding the  mechanisms involved in the production of particles in collisions of heavy nuclei and their subsequent interactions in the medium created in the collision is an important aspect of the physics programs of the Relativistic Heavy Ion Collider (RHIC) and the Large Hadron Collider (LHC). Measurements carried out in the last two decades indicate that a new form of matter, consisting of deconfined quarks and gluons and known as quark–gluon plasma (QGP), is produced in collisions of large nuclei (e.g., Au and Pb) at the very high energies available at these facilities. Evidence for this new form of matter arises in part  from measurements of nuclear modification factors which indicate that the matter produced in these collisions is rather opaque to the propagation of high momentum partons~\cite{Adams:2005dq,Adcox:2004mh,Arsene20051,
Back:2004je,
ALICE:2015mjv,
ALICE:2016flj,
CMS:2017uuv,
CMS:2018zza,
ALICE:2019hno,aliceReview2022,CMS:2012qk,STAR:2017ieb}. Also, observations of collective behavior suggest that the matter formed is strongly interacting and features a nearly vanishing specific shear viscosity~\cite{ALICE:2010suc,ALICE:2013osk,Song:2010mg,Qiu:2011hf,Song:2012tv,Gale:2013da,Bernhard:2019bmu,CMS:2016fnw,ATLAS:2016yzd,CMS:2024krd,STAR:2015wza,PHENIX:2015zbc}.

Recently, there has been great interest in investigating whether such opaque matter can be produced in small collision systems, such as pp and p--Pb collisions. There are a variety of techniques used for such investigations, which include attempts to identify jet quenching relative to the system geometry and efforts to identify collective flow based on multiparticle correlations~\cite{ALICE:2019zfl,ALICE:2014dwt}.
In parallel with these investigations, it is also of interest to determine how the particle production evolves from small to large collision systems~\cite{dnchByDetaVsEnergyPRL16}. Transverse momentum spectra of produced particles are evidently a prime source of such information. However, it is also  found that measurements of differential particle correlations bring additional information for the understanding of the  production of hadrons and their interactions in small and large collision systems. 
Among these, measurements of number ($R_2$) and transverse momentum ($P_2$) two-particle correlation functions have been already explored in p--A and A--A collisions~\cite{AliceDptDptLongPaper,Basu:2020ldt,ALICE:2015nuz}. Furthermore, the measured balance function in pp, p--A, and A--A collisions as a function of multiplicity presents considerable  challenges to the leading models used in heavy-ion physics~\cite{ALICE:2015nuz}.
It is thus of interest to find out whether measurements of these correlation functions in pp collisions can similarly challenge the leading models used towards the description of particle production in  small systems.

 Recent measurements of $\Rtwo$ and $\Ptwo$ have played an important  role in independent verification of the collective nature of azimuthal correlations observed in \PbPb\ collisions~\cite{Adam:2017ucq,Basu:2016ibk}.  Both the $\Rtwo$ and $\Ptwo$ correlation functions are indeed sensitive to the presence  of collectivity  and may contribute to  further elucidation of this phenomenon in small systems relative to that observed in larger systems~\cite{dnchByDetaVsEnergyPRL16,Basu:2020jbk}. 
 Moreover, prior measurements have also  revealed distinct differences in the dependence on \mbox{$\Deta~(\equiv \etaOne - \etaTwo)$} and \mbox{$\Dphi~(\equiv \phiOne - \phiTwo)$}, where $\eta$ and $\varphi$ are pseudorapidity and azimuthal angle of particles 1 and 2, for $\Rtwo$ and $\Ptwo$ correlation functions. The findings indicate that the near-side peak of both charge-independent (CI) and charge-dependent (CD) correlations in $\Ptwo$ is notably narrower than in $\Rtwo$, regardless of the centrality (collision impact parameter)~\cite{ALICE:2013hur} of \PbPb\ collisions~\cite{AliceDptDptLongPaper}. This further supports the idea put forth in Ref.~\cite{Sharma:2008qr} that a comparative analysis of $\Rtwo$ and $\Ptwo$ correlation functions can offer increased sensitivity to the underlying mechanisms governing particle production in small as well as large collision systems.

 The new measurements reported in this work are also designed to enable a better 
understanding of particle production processes underpinning the underlying event of pp collisions, and more specifically the particle production mechanisms involved in soft multiparticle production and the low transverse momentum  components of  jets~\cite{Sahoo:2018uhb}. Measurements in pp are also of interest to study the evolution of these correlation functions with the system size and their compatibility among the different collision systems at similar charged particle multiplicities. Thus, one of the main goals of this work is to provide additional information about the shape and magnitude of the correlation functions in the smallest hadronic  collision systems. Consequently, this study is an important extension of recent measurements of these correlation functions in p--Pb and Pb--Pb collisions by the ALICE Collaboration~\cite{AliceDptDptLongPaper}. The results  involve measurements of the  $R_2$ and $P_2$ correlation functions, and their characteristics, for CI and CD combinations of charged particles. These are compared with PYTHIA8 predictions to verify whether this model is capable of providing a reasonable description of correlated particle production in small collision systems. Furthermore, inclusive charge balance function in pp collisions was measured. Charge balance functions have been exploited primarily in collisions of heavy nuclei to identify the presence of an extended period of isentropic expansion in these systems~\cite{ALICE:2015nuz,CMS:2023sua}, but recent theoretical works also indicate that they enable the estimation of the diffusivity of light quarks and may also serve in the determination of QGP susceptibilities~\cite{Pratt:2015jsa,Pratt:2019pnd}.

This paper presents a comparative analysis of $\Rtwo$ and $\Ptwo$  correlation functions  measured in
pp collisions at a center-of-mass energy $\s$ = 13 TeV. As in prior analyses of p--Pb and Pb--Pb collisions~\cite{Adam:2017ucq,AliceDptDptLongPaper}, the two correlation functions are first measured for like-sign (LS) and unlike-sign (US) charged particle pairs. These are then combined to CI and CD correlation functions, as described in Sec.~\ref{sec:def}. These correlations are characterized by their azimuthal and longitudinal widths and compared with characteristics of $\Rtwo$ and $\Ptwo$ correlation functions measured in larger collision systems.

The article is organized as follows. The correlation functions, $\Rtwo$
and $\Ptwo$, and their LS, US, CI, and CD components are defined in
Sec.~\ref{sec:def}. Section~\ref{sec:dataAnalysis} presents a  summary of the data taking conditions as well as the various technical details of the analysis, including event and  track
selection criteria, efficiency correction, quality tests, etc. The basic configuration of the Monte Carlo (MC) model used for quality control and towards the interpretation of the measured data  is briefly described in Sec.~\ref{sec:mc}. 
Section~\ref{sec:errors} presents a discussion of the techniques used for the estimation of the  statistical and systematic uncertainties.  The experimental results are
presented in Sec.~\ref{sec:results} where they are compared with PYTHIA8 predictions. A summary is provided in Sec.~\ref{sec:summary}.

\section{Definitions of observables}
\label{sec:def}
Particle number and  transverse momentum correlations  
are reported based on $\Rtwo$ and $\Ptwo$~\cite{Sharma:2008qr,Sahoo:2018uhb}  defined in terms of
single- ($\rhoOne$) and two-particle ($\rhoTwo$) densities 
\be \label{eq:densities}
\rhoOne (\eta_{\mathrm{j}}, \varphi_{\mathrm{j}}) &=& \frac{\mathrm{d}^2 \it{N}}{\deta_{\mathrm{j}} \dphi_{\mathrm{j}}}, \\ 
\rhoTwo(\etaOne, \phiOne,\etaTwo, \phiTwo)&
=&  \frac{\mathrm{d}^4 \it{N}}{\detaOne\dphiOne\detaTwo \dphiTwo}, 
\ee
where $\eta_{j}$, $\varphi_{j}$ ($j=1,2$) are the pseudorapidities and 
 azimuthal angles  of particles 1 and 2, respectively.

The number correlation function, $\Rtwo$, is formulated as a two-particle cumulant normalized by the product of single-particle densities according to
\be
\label{eq:R2}
\Rtwo (\etaOne, \phiOne,\etaTwo, \phiTwo)&
=& \frac{\rhoTwo(\etaOne, \phiOne,\etaTwo, \phiTwo) }{ \rhoOne 
 (\etaOne, \phiOne)\rhoOne (\etaTwo, \phiTwo)}-1,
\ee
whereas the  dimensionless transverse momentum correlation function, $P_2$, is defined as the ratio between  $\la \Delta \pt\Delta \pt \ra$  and  the square of the mean transverse momentum, $\la \pt \ra$. This can be expressed as follows
\be
\label{eq:P2}
\Ptwo (\etaOne, \phiOne,\etaTwo, \phiTwo)&
=&\frac{\DeltaPtPt(\etaOne, \phiOne,\eta_2, \phiTwo)}
{\la \pt\ra^2}.
\ee
The $\la \Delta \pt\Delta \pt \ra$ differential correlation function is defined as
\begin{equation}\label{eq:dptdpt}
\la \Delta \pt\Delta \pt \ra(\etaOne, \phiOne,\eta_2, \phiTwo) = 
\frac{  \int_{p_{\mathrm T,\min}}^{p_{\mathrm T,\max}} \Delta p_{\mathrm T,1} 
 \Delta p_{\mathrm T,2}  \  \rho^{'}_2({\vec p}_{1},{\vec p}_{2}) \ {\mathrm d}p_{\mathrm T,1}{\mathrm d}p_{\mathrm T,2}  
  } 
{\int_{p_{\mathrm T,\min}}^{p_{\mathrm T,\max}} \rho^{'}_2({\vec p}_{1},{\vec p}_{2}) \ {\mathrm d}p_{\mathrm T,1}{\mathrm d}p_{\mathrm T,2} },
\end{equation}
in which $\rho^{'}_2({\vec p}_{1},{\vec p}_{2})$ is analogous for $\rho_2$, but is expressed as a function of the momenta ${\vec p}_{1}$ and ${\vec p}_{2}$ of the particles constituting the pair instead of their $\eta$ and $\varphi$. Here, $p_{\mathrm T,\min}$ and $p_{\mathrm T,\max}$ specify the transverse momentum range of the measurement. The quantities $\Delta\pti = \pti - \la \pt \ra$, where $i=1,2$, are deviations from the average transverse momentum  calculated as follows
\be 
\la \pt \ra = \int_{p_{\mathrm T,\min}}^{p_{\mathrm T,\max}}\rhoOne \pt \dpt / \int_{p_{\mathrm T,\min}}^{p_{\mathrm T,\max}}\rhoOne \dpt.
\ee

By construction, both $R_2$ and $P_2$ are robust observables to first order. Their magnitude remains insensitive to particle losses (i.e., detection inefficiencies) provided that these inefficiencies exhibit only modest dependence on kinematic variables, assuming the efficiency is uniform across the $\pt$ acceptance of the measurement. Furthermore, both observables are dimensionless and their magnitude can be used as reliable measure of the degree of correlation between the produced particles. However, $\Ptwo$ explicitly incorporates deviations of particle momenta to the mean  and is therefore sensitive to  the ``hardness" of the correlations. This means it can distinguish whether correlated particle pairs involve soft--soft or hard--hard interactions, i.e., both particles below $\la \pt \ra$ or both above $\la \pt \ra$. It can also identify soft--hard pairs, where one particle has a $p_{\rm T}$ below the mean and the other above. Additionally, it is important to note that the relative magnitudes of contributions from soft--soft, hard--hard, and soft--hard combinations may change as a function of the ($\Deta, \Dphi$) pair separation. One expects, for instance, that particle correlations within jets should yield a preponderance of hard–hard correlations near the core (thrust axis) of a jet. However, soft–hard dominance is expected for pairs involving one particle near the thrust axis and one emitted at a large angle relative to that axis. This implies that $P_2$ features an added sensitivity to the angular ordering of particle production in jets as well as in resonance decays~\cite{Sahoo:2018uhb}. The scaling properties of $R_2$ and $P_2$  with system size have been described in Ref.~\cite{Sharma:2008qr}. 

In this work, the correlation functions $\Rtwo$ and $\Ptwo$ are reported 
as functions of $\Deta$ and $\Dphi$ by averaging their  magnitude  across the pair average pseudorapidity 
$\bar \eta \equiv \frac{1}{2}(\etaOne + \etaTwo)$  and  average azimuthal angle $\bar \varphi \equiv \frac{1}{2}(\phiOne + \phiTwo)$ acceptance,
according to
\be \label{eq:fact}
O(\Deta, \Dphi) 
= \frac{1}{\Omega(\Deta)} \int_{\Omega} \mathrm{O}(\Delta\eta,\bar \eta,\Delta\varphi,\bar \varphi) {\mathrm d}\bar\eta {\mathrm d}\bar\varphi .
\ee
Here, $O(\Deta, \Dphi)$ is either $\Rtwo(\Deta, \Dphi)$ or $\Ptwo(\Deta, \Dphi)$. The variable $\Omega(\Deta)$, which depends solely on the $\Deta$, represents the width of the acceptance in
$\bar{\eta}$ at a given value of $\Deta$ and $\Dphi$~\cite{Pruneau_2017}. Furthermore, $\Rtwo$ and $\Ptwo$  are  determined for $\Delta\varphi$  modulo $2\pi$ and shifted by $-\pi/2$ for convenience of representation in the figures.

The measured densities $\rhoOne$ and $\rhoTwo$ are  directly impacted by  detection inefficiencies. Given that these differ for positively and negatively charged particles at  given values of $\eta$ and $\varphi$,  benefiting from the  robustness of $R_2$ and $P_2$ correlation functions requires these to be measured independently for pairs of $(+,+)$, $(-,-)$, $(-+)$, and $(+,-)$ charged particles.  Correlation functions
for $(+-)$ and  $(-+)$ pairs are then averaged to yield US
correlation functions, $O^{\mathrm US}\equiv \frac{1}{2}[ {O}^{+-} + {O}^{-+}]$, and
correlation functions of pairs  $(++)$ and $(--)$ are averaged to yield LS
correlation functions,  $O^{\mathrm LS} \equiv \frac{1}{2}[ {O}^{--} + {O}^{++}]$.
In turn, $O^{\mathrm US}$ and $O^{\mathrm LS}$ correlation functions are combined
into CI and CD correlation functions according to

\begin{align} 
\label{eq:CI}
O^{\mathrm{CI}} &= \frac{1}{2}\left[O^{\mathrm{US}} + O^{\mathrm{LS}}\right],\\
\label{eq:CD}
O^{\mathrm{CD}} &= \frac{1}{2} \left[O^{\mathrm{US}} - O^{\mathrm{LS}}\right].
\end{align} 
The CI functions measure the average of correlations between all charged particles, whereas the CD observables are sensitive to the difference of US and
LS pairs and are thus largely driven by charge conservation effects. 

The width $\sigma_{\Omega}$ of the near-side peak of $\Rtwo$ and $\Ptwo$ correlation functions of CI and CD pair combinations is 
calculated along the $\Deta$ and $\Dphi$ axes with the procedure already used in prior ALICE studies~\cite{AliceDptDptLongPaper}
\be
\label{eq:widths}
\sigma_{\Omega} = \left( \frac{\sum_{i}[O(\Omega_{i}) -
    T]\Omega_{i}^{2}} {\sum_{i}[O(\Omega_{i}) - T]}  \right)^{1/2},
\ee

where $i$ iterates over the bins, and $T$ is an offset or threshold value. Offsets 
are considered to prevent width values that are determined simply by the detector acceptance. In order to calculate the width along $\Dphi$, offsets are estimated by averaging three narrow $\Dphi$ intervals near the minimum of the $\Dphi$ distribution. On the other hand, while evaluating the width along $\Deta$, offsets are determined close to the edge of the acceptance, approximately  at $\Deta \approx 1.6$. Since the correlation vanishes for large $|\Deta|$ values  in the case of $\RtwoCD$, a null offset is used, which results in the exclusion of contributions from the unobserved part beyond the acceptance~\cite{AliceDptDptLongPaper}.

The  balance function ($B$)~\cite{S.PrattPRL:2000BalFun1st}  of charged particles is also measured in minimum bias pp collisions. It is computed according to 
\be
\label{eq:balanceFun}
{B}\DetaDphi \equiv \Big\langle \frac{\mathrm{d^{2}}N_\mathrm{ch}}{\mathrm{d}\eta
  d\varphi} \Big\rangle \times \RtwoCD\DetaDphi,
\ee 
which applies when densities (yields) of positively and negatively charged particles are approximately equal~\cite{C.PruneauPRC:2002Fluct,ALICE:2015ial}. In addition, the integral of the charge balance function, $I_{\mathrm{B}}$,  is calculated according to 
\be
\label{eq:IB}
I_{\mathrm{B}} = {\int_{\mathrm{d}\eta_{\min}}^{\mathrm{d}\eta_{\max}}\int_{\mathrm{d}\varphi_{\min}}^{\mathrm{d}\varphi_{\max}} \int_{\mathrm{d}p_{\mathrm T,\min}}^{\mathrm{d}p_{\mathrm T,\max}}} B {\DetaDphi} {\mathrm{d}\eta \mathrm{d}\varphi \mathrm{d}p_{\mathrm{T}}}.
\ee 

The integral of $B$ in pp collisions is compared with previously published results in \PbPb~collisions~\cite{genBF}. This comparative study provides insights into how the charges are balanced and produced in both  pp and \PbPb~collisions.

\section{Datasets and experimental method}
\label{sec:dataAnalysis}

Results reported in this work are based on an analysis of $4.4\times 10^8$ minimum bias (MB) pp collisions at $\s$ = 13 TeV collected  by the ALICE detector during the  Run 2 data taking campaign in 2018. The MB trigger selects collisions with at least one hit in both the V0A and V0C detectors, which are scintillator arrays covering the pseudorapidity ranges  2.8 $< \eta <$ 5.1 and $-3.7 < \eta < -1.7$, respectively. Moreover, to minimize instrumental effects and maintain  approximately uniform acceptance and efficiency as a function of pseudorapidity, only events having a reconstructed primary vertex (PV) within $\pm 8$ cm from the nominal center of the ALICE detector along
the beam direction are considered in the analysis. Pile-up events involving multiple reconstructed vertices are rejected using offline algorithms~\cite{alicePerf}. Charged particle tracks included in this analysis were reconstructed  using the Inner Tracking System (ITS) and the Time Projection Chamber (TPC) detectors. The design and performance of the V0, ITS, and TPC detectors are reported in Refs.~\cite{ALICE:2008ngc,v0Det,tpcDet}.

The analysis is limited to charged-particle tracks reconstructed  within the pseudorapidity range \mbox{$|\eta | < $ 0.8} and the transverse momentum interval $0.2 < \pt < 2.0$ GeV/$c$
to emphasize particle production governed by non-perturbative soft quantum chromodynamics (QCD) processes. The analysis  includes selection criteria to suppress secondary charged particles (i.e., particles originating from weak decays, $\gamma$-conversions, and secondary hadronic interactions in the detector material) and fake tracks (random associations of space points). The used track parameters are obtained by the Kalman filter  at the collision primary vertex. Tracks of particles originating from weak decays of K$^{0}_{\mathrm S}$ and
$\Lambda^{0}$ and other secondaries are suppressed based on a
$\pt$-dependent selection on the distance of closest approach
(DCA) of charged particle trajectories to the PV~\cite{ALICE:2015olq}. Electrons and positrons are rejected based on their  specific  energy loss (\dedx) measured within the TPC.
Good track quality is
assured by retaining only tracks with more than 70 reconstructed TPC
space points, out of a maximum of 159 and a momentum fit with a $\chi^{2}$-value per degree of freedom less than 2.0. In order to increase the track quality and further suppress tracks produced in pile-up of collisions occurring within the long TPC readout time, the selected tracks are required to have a combined refit in both the ITS and the TPC and at least one hit in the innermost part of the ITS.

Equations ~\ref{eq:densities}--~\ref{eq:fact} are used to obtain the $R_2$ and $P_2$ correlation functions as  discussed in the previous section. Given that the accuracy of $R_2$ and $P_2$  may be impacted by  $p_{\mathrm T}$-dependent inefficiencies, both $R_2$ and $P_2$ correlation functions are explicitly corrected for such dependencies~\cite{Sharma:2008qr}. The efficiencies are determined using Monte Carlo simulations of pp collisions based on the PYTHIA8~\cite{Skands:2014pea} event generator and the GEANT4~\cite{GEANT} transport code, as explained in detail in Sec.~\ref{sec:mc}.


\section{Monte Carlo model studies}
\label{sec:mc}
Monte Carlo (MC) simulations of pp collisions at $\s$ = 13 TeV generated with PYTHIA8~\cite{Skands:2014pea} are used to determine efficiency correction factors, evaluate the performance of the analysis procedure (also known as closure test), and produce the correlation functions compared with the experimental data in Sec.~\ref{sec:results}. The PYTHIA8  event generator is  based on a QCD  description of quark and gluon interactions    at the leading order (LO) and uses the Lund string fragmentation model~\cite{lundFrag} for 
high-$p_{\rm T}$ parton hadronization. The production of soft particles (i.e., the underlying event) is handled through fragmentation of mini-jets from initial and final state radiation, as well as multiple parton interactions~\cite{Sjostrand:2007gs}. The calculations are performed with the Monash 2013 tune of PYTHIA8 running in minimum-bias mode, with soft QCD processes and color reconnection turned on~\cite{Skands:2014pea}.

Single particle detection efficiencies are estimated based on the ratio of detector and generator levels single particle yields obtained with PYTHIA8. Reconstructed (REC) level yields are obtained by propagating PYTHIA8  events  through a model of the ALICE detector with GEANT4 and reconstructing the simulated events with the same software used for real data. Generator (GEN) level yields are obtained directly from the primary particles produced by PYTHIA8, without including efficiency losses or resolution smearing. Subsequently, tracking efficiency corrections are applied to both single- and two- particle pairs; these corrections cancel to first order in the calculations of $\Rtwo$ and $\Ptwo$.

A MC closure test is also carried out based on simulated PYTHIA8 events processed with GEANT4 and the full ALICE reconstruction of $R_2$ and $P_2$ correlation functions. The REC calculations are compared with correlation functions obtained at the GEN level to identify possible biases in the experimental determination of these measurements.  Differences between  reconstructed and generator level  $R_2$ and $P_2$ correlation functions are found to be of the order or smaller than 1$\%$. These differences, though modest, are conservatively added in the evaluation of   systematic uncertainties discussed in the next section. 

Furthermore, small particle losses are caused by momentum resolution  near the edges of the acceptance: when charged particles from the generator level pass through GEANT4, a slight shift in $\eta$ and $\pt$ due to resolution effects could move these particles outside of the acceptance, leading to particle loss. This was addressed by excluding particles from the MC model that fell outside the acceptance and comparing the results with those obtained using the default approach, in which those particles were retained. The resulting difference was assigned as a part of the systematic uncertainty.

\section{Determination of statistical and systematic uncertainties}
\label{sec:errors}
Given the fact that the statistical uncertainties of the  correlation functions are highly correlated, the final statistical uncertainties are estimated with the sub-sample method~\cite{Bevington:1305448}. Overall, eleven data samples, obtained by splitting the full data sample collected during the 2018 period, are used in the analysis separately. Their weighted mean constitutes the final result and  the standard deviations from the mean are used  to estimate 
the  statistical uncertainties on the amplitude of the correlation functions.

Systematic uncertainties are estimated by repeating the analysis
with modified event and charged-track selection criteria for LS, US, CI, and CD pairs separately. The criteria are  varied individually to assess their impact on the measured correlation functions and their characteristics. As discussed in Sec.~\ref{sec:mc}, a correction for the tracking efficiency is applied in the default analysis procedure. To estimate the systematic uncertainty due to possible imperfections in the description of the tracking efficiency, the analysis was repeated without applying this correction. A similar approach is applied to estimate the systematic uncertainty due to the correction for particle loss. As a cross check that no bias is introduced by the pile-up rejection at the event and track selection levels, the analysis was repeated without pile-up removal and a negligible effect on the results was found.
\begin{table}[tb]
  \caption{Sources of systematic uncertainties.}

  \centering
  \begin{tabular}{ccccccc}
    \hline
    \hline
    Parameter
    & Default
    & Variation\\
    \hline    
    DCA$_{\mathrm {xy,z}} $(cm)
    & 0.2
    & 0.1\\

    Track selection
    & Global tracks
    & tracks reconstructed based only on TPC information\\

    Tracking efficiency
    & With
    & Without\\

    Magnetic field polarity (T)
    & +0.5
    & -0.5\\

    No. of TPC space points
    & $>$70
    & $>$90\\

    Particle loss
    & With
    & Without\\

    Track pile-up
    & Removed 
    & Included\\

    Event pile-up
   & Removed 
    & Included\\
    \hline
    \hline
  \end{tabular}
    \label{tab:DifftSysSource}   
\end{table}
The different selection criteria used to estimate the systematic uncertainties are described below and listed in Tab.~\ref{tab:DifftSysSource}. The Barlow  criterion~\cite{barlow} is
used to assess the statistical significance of differences observed when changing the selection criteria relative to the nominal analysis. Total systematic uncertainties are estimated by assuming that the different sources are   independent: contributions from the various sources are summed in quadrature to obtain the total systematic  uncertainties on the amplitude of the $R_2$ and $P_2$ correlation functions, as well as quantities derived from these. The maximum contributions of systematic uncertainties and their sources are presented in Tab.~\ref{tab:maxSysProj} and~\ref{tab:maxSysWidth} for the projections and widths of the correlation functions, respectively.

\begin{figure}[tb]
  \centering
  \includegraphics[scale=.37]{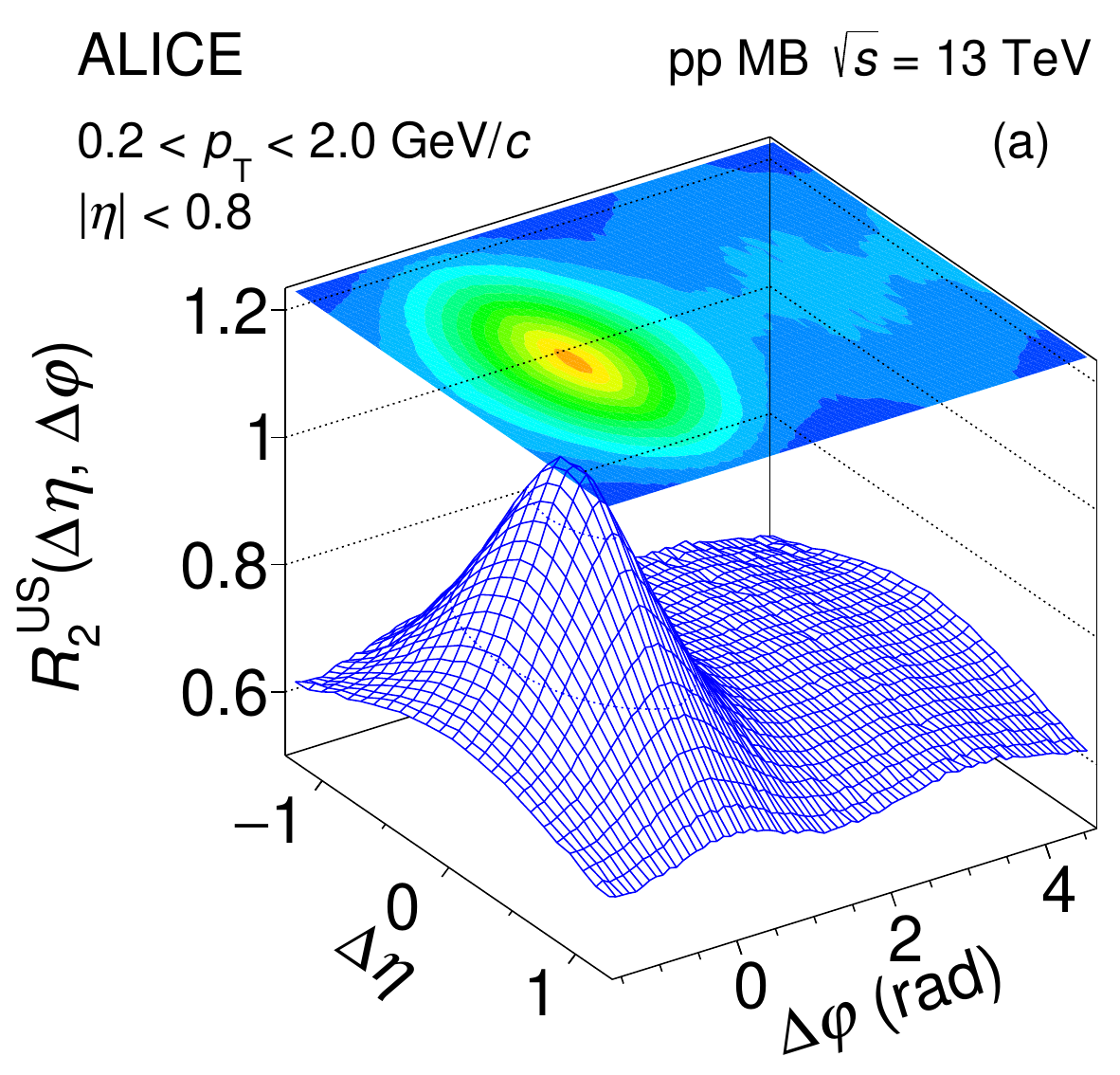}
  \hspace{0.5cm}
  \vspace{0.5cm}
  \includegraphics[scale=.37]{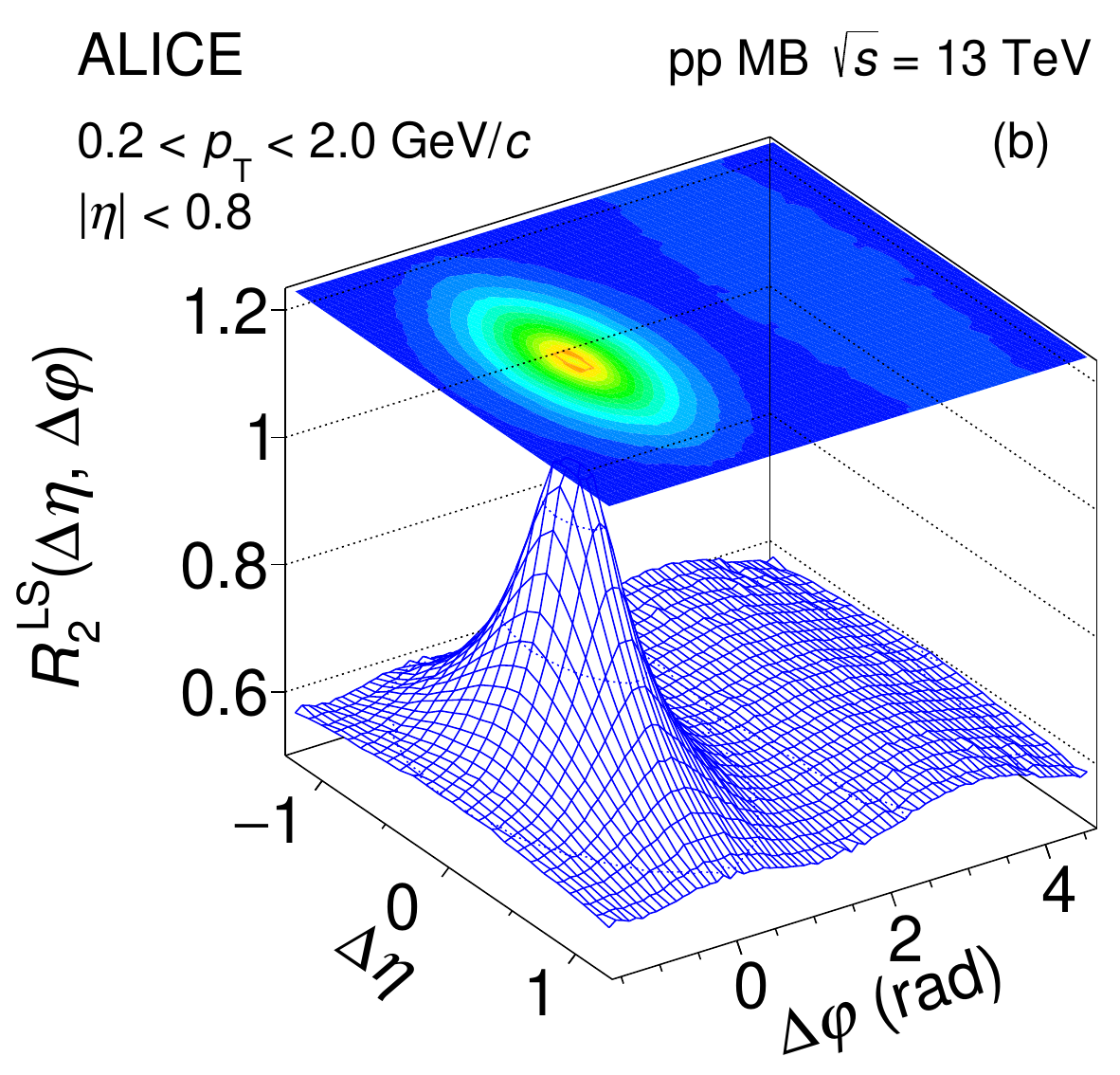}
  \includegraphics[scale=.37]{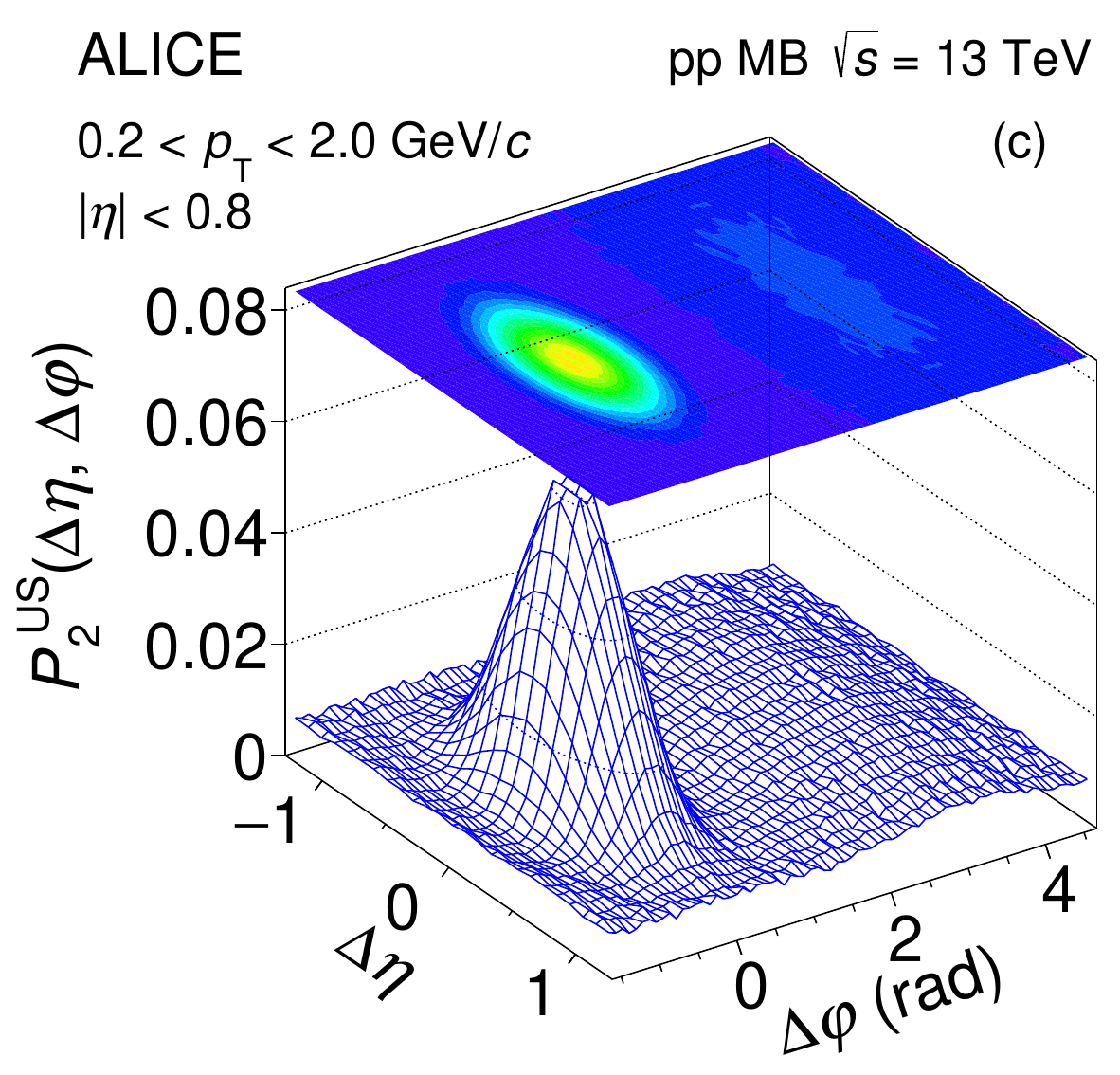}
  \hspace{0.5cm}
  \vspace{0.5cm}
  \includegraphics[scale=.37]{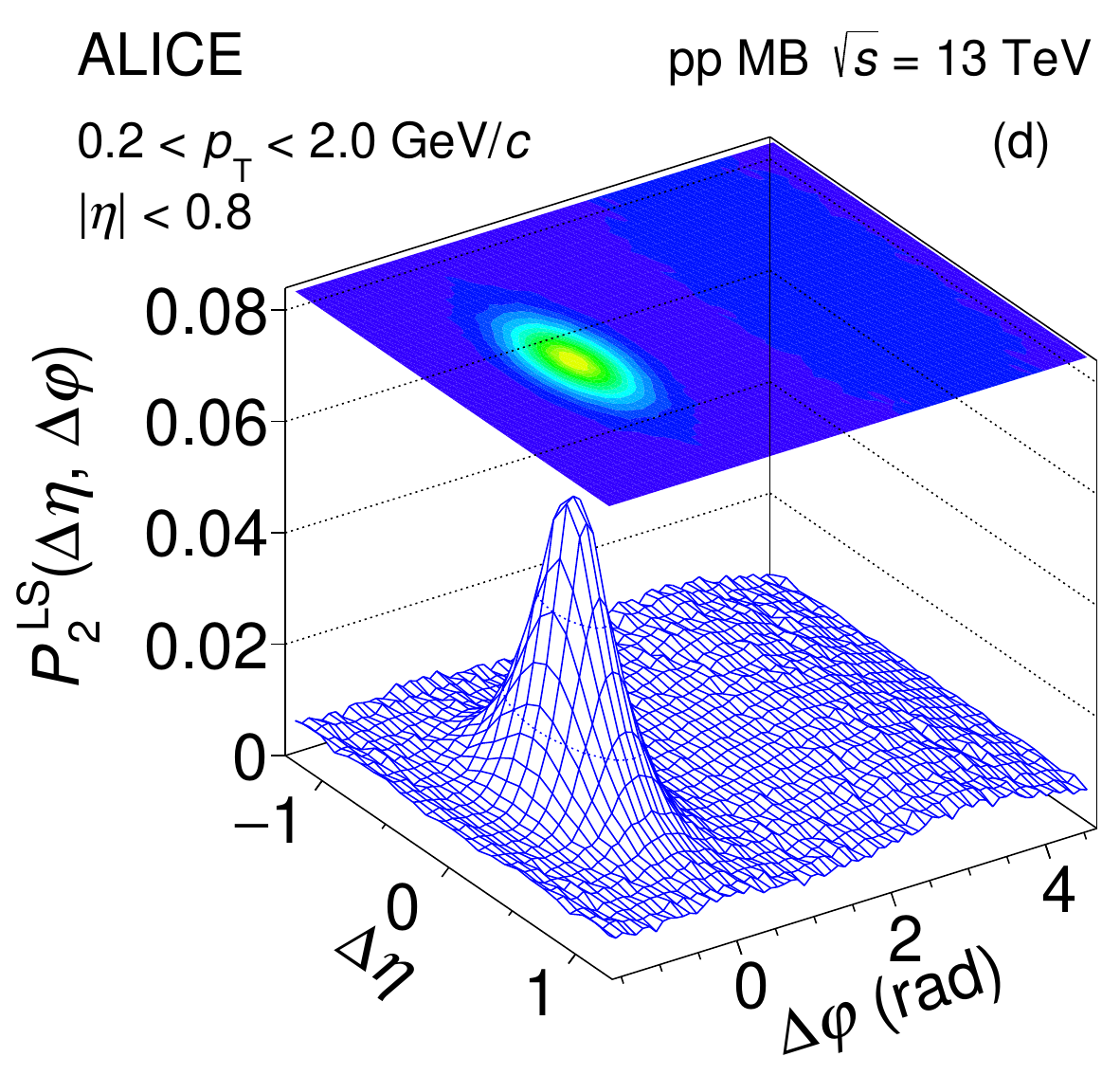}
 \caption{Correlation functions (a) $\RtwoUS$, (b) $\RtwoLS$, (c) $\PtwoUS$, (d) $\PtwoLS$ of charged hadrons measured in minimum bias pp collisions at $\s$ = 13 TeV. Charged hadrons are selected in the  range $0.2 < \pt < 2.0$ GeV/$c$ and with pseudorapidity $|\eta| < 0.8$.}
 \label{fig:r2p2UsLs}
\end{figure}

\begin{table}[tb]
  \centering
    \caption{Maximum systematic uncertainties of the correlation functions and their sources for projections.}
\begin{tabular}{cccc}
    \hline
\hline    
    Correlator
    & Category
    & \Centerstack{ Source} & \Centerstack{ Maximum systematic uncertainties ($\%$)}\\
    \hline
$\RtwoCI$
& \Centerstack{ $\Deta$ \\ $\Dphi$ }
    & \Centerstack{ Tracking efficiency\\ Tracking efficiency}
    & \Centerstack{ 0.19\\ 0.13}\\

$\PtwoCI$
& \Centerstack{ $\Deta$ \\ $\Dphi$ }
    & \Centerstack{ MC closure test\\ MC closure test}
    & \Centerstack{ 0.04\\ 0.03}\\
$\RtwoCD$
  & \Centerstack{ $\Deta$ \\ $\Dphi$ }
    & \Centerstack{ MC closure test\\ MC closure test}
    & \Centerstack{ 0.04\\ 0.03}\\

$\PtwoCD$
& \Centerstack{ $\Deta$ \\ $\Dphi$ }
    & \Centerstack{ DCA$_{\mathrm {xy,z}} $\\ No. of TPC spcae points}
    & \Centerstack{ 0.0008\\ 0.0013}\\
\hline
 \hline
  \end{tabular}    \label{tab:maxSysProj}   
\end{table}


\begin{table}[tb]
  \centering
    \caption{Maximum systematic uncertainties of the correlation functions and their sources for widths.}
\begin{tabular}{cccc}
    \hline
\hline    
    Correlator
    & Category
    & \Centerstack{ Source} & \Centerstack{ Maximum systematic uncertainties ($\%$)}\\
    \hline
$\RtwoCI$
& \Centerstack{ $\Deta$ \\ $\Dphi$ }
    & \Centerstack{ DCA$_{\mathrm {xy,z}} $\\ Track pile-up}
    & \Centerstack{ 0.36\\ 0.58}\\

$\PtwoCI$
& \Centerstack{ $\Deta$ \\ $\Dphi$ }
    & \Centerstack{ DCA$_{\mathrm {xy,z}} $\\ Tracking efficiency}
    & \Centerstack{ 0.59\\ 0.27}\\

$\RtwoCD$
  & \Centerstack{ $\Deta$ \\ $\Dphi$ }
    & \Centerstack{ Tracking efficiency\\ Tracking efficiency}
    & \Centerstack{ 0.46\\ 2.17}\\

$\PtwoCD$
& \Centerstack{ $\Deta$ \\ $\Dphi$ }
    & \Centerstack{ Tracking efficiency\\ No. of TPC spcae points}
    & \Centerstack{ 0.54\\ 1.77}\\
\hline
 \hline
  \end{tabular}    \label{tab:maxSysWidth}   
\end{table}

Tracking efficiency contributes the most to the systematic uncertainties
for projections of $\RtwoCI$ along $\Deta$ ($\Dphi$), which  is
roughly 0.19$\%$ (0.13$\%$). Similarly,  the highest contribution to
the systematic uncertainties of the projections of $\RtwoCD$ and
$\PtwoCI$ is 0.04$\%$ (0.03$\%$) along $\Deta$ ($\Dphi$), resulting
from Monte Carlo closure tests. The largest contribution to the
systematic uncertainties for projections of $\PtwoCD$ along $\Deta$
($\Dphi$) is  0.0008$\%$ (0.0013$\%$), which is due to DCA (impact of
the TPC sector boundaries). Contributions to the systematic uncertainties
from other sources are found to be negligible.


The systematic uncertainties on  the widths of the correlation functions, presented in  Sec.~\ref{sec:width}, were assessed in a similar fashion by varying selection criteria individually. Contamination from secondaries not rejected by the DCA selection criteria  contribute the most to these  uncertainties, approximately 0.36$\%$ (0.59$\%$),
on the width of $\RtwoCI$ ($\PtwoCI$) along $\Deta$. The largest contribution to the systematic uncertainties in the width of $\RtwoCI$ ($\PtwoCI$) as a function of $\Dphi$ arises from track pile-up effects (tracking efficiency), amounting to approximately 0.58$\%$ (0.27$\%$). However, the highest contribution to
the systematic uncertainties on the width of $\RtwoCD$ along $\Deta$
($\Dphi$) is 0.46$\%$ (2.17$\%$) and originates from uncertainties on track
reconstruction efficiencies. Tracking efficiencies (impact of
the TPC sector boundaries) give rise the biggest contribution to the systematic uncertainties for the width of $\PtwoCD$ along $\Deta$
($\Dphi$), which is around 0.54$\%$ (1.77$\%$). Contributions from other sources are insignificant in this case.

The systematic uncertainties on the single particle density, shown in Sec.~\ref{sec:bf}, contribute the most to the systematic uncertainties on the magnitude of the balance function and its integral. Altogether, the systematic uncertainty on the integral amounts to  5.8$\%$, with the contribution due to single particle density accounting for about 5.7$\%$.

\section{Results}
\label{sec:results}

The $R_2$ and $P_2$ correlation functions were first determined in two dimensions, i.e., as functions of particle pair pseudorapidity difference ($\Delta\eta$) and azimuthal angle difference ($\Delta\varphi$). These were then projected onto $\Delta\eta$ and $\Delta\varphi$ axes to further examine the dependence  of the correlation functions on these kinematic variables.

\subsection{Charge combinations of correlation functions}
\label{sec:r2p2basic}
\begin{figure}[tb]
  \includegraphics[scale=0.36]{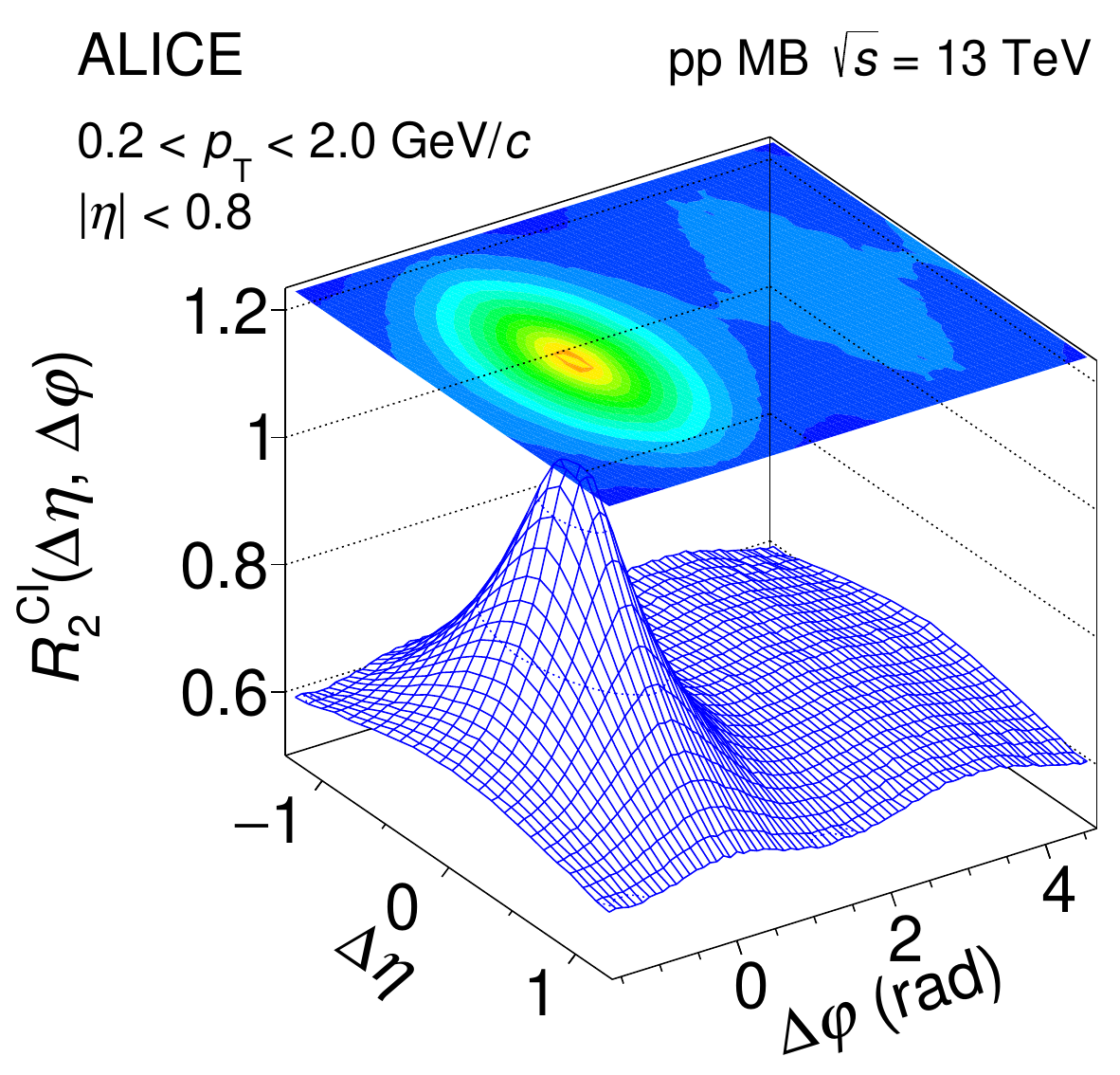}
   \qquad
  \includegraphics[scale=0.36]{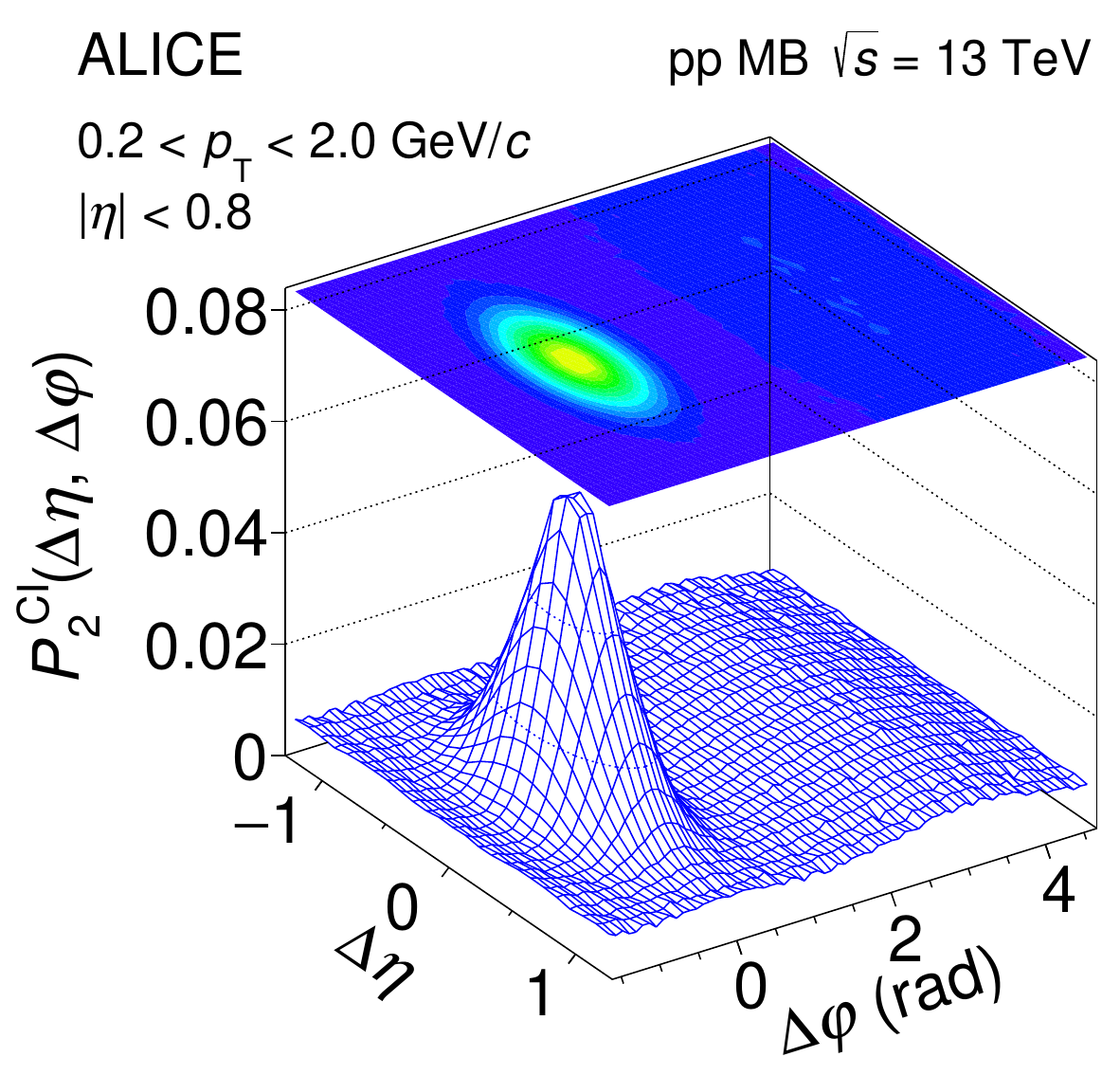}   
  \caption{Correlation functions $\RtwoCI$ (left) and $\PtwoCI$
    (right) of charged hadrons measured in minimum bias 
    pp collisions at $\s$ = 13 TeV. Charged hadrons are selected in the  range $0.2 < \pt < 2.0$ GeV/$c$ and with pseudorapidity $|\eta| < 0.8$.}
  \label{fig:r2p2CiSys}%
\end{figure}

The $\Rtwo$ and $P_2$ correlation functions for US and LS charged-particle pairs forming the basis of the measurements are displayed in
Fig.~\ref{fig:r2p2UsLs}. These correlation functions share common features, but also exhibit significant differences. All four correlation functions are dominated by the presence of a strong and relatively narrow peak centered at ($\Delta\eta, \Delta\varphi) = (0,0)$, hereafter called the near-side peak because it corresponds to the emission of two particles near one another  in $\Dphi$. The correlation functions also feature extended structures, of a smaller amplitude, hereafter called away-side, because they correspond to particle emission in two different hemispheres, at $\Dphi \approx $ 180$^\circ$. However, it is important to note also that US correlation functions feature a wider peak on the near-side as shown in Figs.~\ref{fig:r2p2UsLs} (a) and (c), whereas the LS ones exhibit a relatively narrower near-side peak, as shown in Figs.~\ref{fig:r2p2UsLs} (b) and (d). The narrower shape of the near-side peak of the LS correlation functions arises, in part, from Bose--Einstein quantum statistics (also known as  Hanbury Brown--Twiss (HBT) effect): two  identical   mesons (e.g., $\pi^{+}\pi^{+}$) are likely to be emitted in nearly the same direction with similar transverse momenta. The longitudinal and azimuthal widths of the LS near-side peak shall thus be partially driven by the inverse of the system size 
(in spatial coordinates)~\cite{ALICE:2023sjd}. Moreover, it can be observed that away-side structures have slightly  different shapes and dependence on $\Delta\eta$. These near- and away-side shapes have formerly been observed in measurements of $C_{2}$ correlation functions (involving a particle with higher $\pt$, known as trigger particle and a second particle considered as associate)   measured in several collision
systems and a variety of beam energies~\cite{Adam:2017ucq,AliceDptDptLongPaper,PhysRevC.77.011901,PhysRevC.75.034901,Abelev:2009af,Adare:2008ae}. Several  different
mechanisms, including dijet, resonance decays, Bose--Einstein
correlations, collective flow, and conservation  of energy and momentum, are thought to be  responsible
for the near- and away-side
shapes~\cite{trigAsso1,trigAsso2,trigAsso3}. Furthermore, it can be observed that the US and LS combinations of $\Ptwo$ correlation functions
exhibit similar behaviors, but significantly  narrower near-side peaks as compared to $\Rtwo$, as seen in Figs.~\ref{fig:r2p2UsLs} (c) and (d). The $\Rtwo$ and $\Ptwo$ US and LS correlation functions, shown in Fig.~\ref{fig:r2p2UsLs},
are combined using Eqs.~\ref{eq:CI} and ~\ref{eq:CD}, respectively. This combination produces the CI and CD of $\Rtwo$ and $\Ptwo$
correlation functions presented in Figs.~\ref{fig:r2p2CiSys} and~\ref{fig:r2p2CdSys}. 

\begin{figure}[tb]
  \centering
  \includegraphics[scale=0.7]{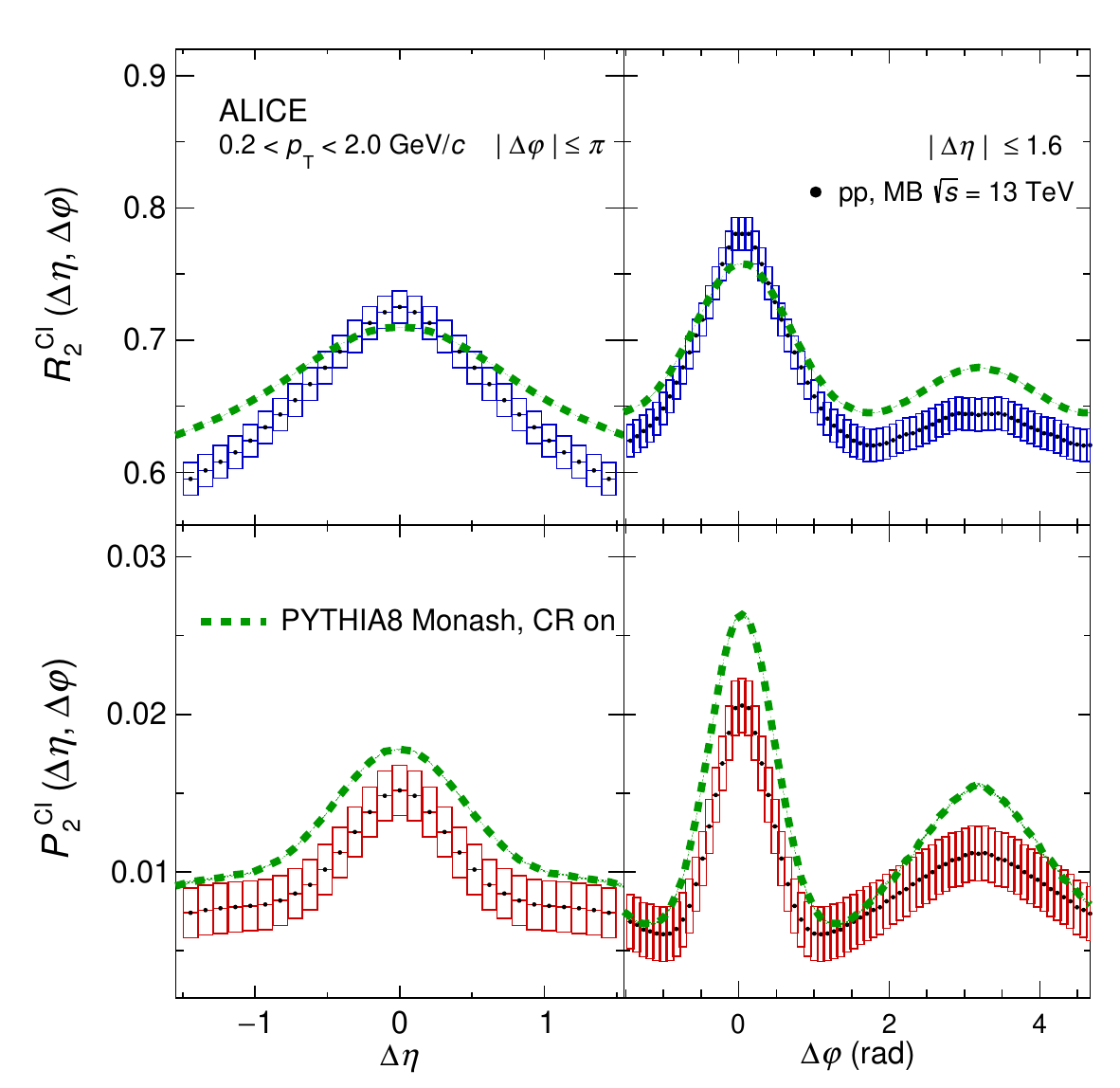}   
  \caption{Projections onto $\Deta$ (left column) and $\Dphi$ (right column) of $\RtwoCI$ (top row)
  and $\PtwoCI$ (bottom row) correlation functions of
  charged hadrons calculated in the selected $\pt $ range
  in pp collisions at \mbox{$\s$ = 13  TeV}. The $\Deta$ and $\Dphi$ projections are calculated as averages of the two-dimensional 
  correlations in the range $|\Dphi| \leq \pi$ and $|\Deta| \leq 1.6$, respectively. Vertical bars and
boxes represent statistical and systematic uncertainties,
respectively. Simulations using PYTHIA8 with the Monash 2013 tune and color reconnection (CR) enabled, as described in the text, are shown as green dashed lines.}
 \label{fig:r2p2CiProjData}
\end{figure} 

\subsection{Charge-Independent (CI) correlation functions}
\label{sec:CI}
The $\Rtwo$ and $\Ptwo$ correlation functions for CI combinations, shown in the left and right panels of Fig.~\ref{fig:r2p2CiSys}, are obtained by averaging US and LS correlation functions. Hence, they evidently feature rather similar dependencies on $\Delta\eta$ and $\Delta\varphi$.  Indeed, both correlation functions have a strong
near-side peak centered at $(\Deta, \Dphi) = (0,0)$ and an away-side
structure centered at $\Dphi = \pi$ and extending over the whole $|\Deta|$ range. It can be seen, however, that the near-side peak of   $P_2^{\rm CI}$ is significantly narrower than the near-side peak of  $R_2^{\rm CI}$. This is better visible in Fig.~\ref{fig:r2p2CiProjData} which  presents
the projections of $\RtwoCI$ and $\PtwoCI$ onto $\Deta$ and $\Dphi$ axes in
the left and right panels, respectively. It is clear that the near-side peak of
the $\PtwoCI(\Deta)$ correlation function is narrower than the corresponding
peak of $\RtwoCI(\Deta)$. It is worth stressing, however, that $R_2$
and $P_2$ correlation functions have the same kernel, i.e., the same source of correlated particles. The fact that the
$\PtwoCI$ near-side peak is narrower relative to that of $\RtwoCI$ is thus
associated with the $\Delta \pt\Delta\pt$ dependence of the former: at
some large relative angles and pseudorapidity differences, the factor $\Delta
\pt\Delta \pt$ manifestly suppresses the amplitude of the $P_2$
correlation function. In the tail of the near-side peak, the contributions of particle pairs with positive and negative values of $\Delta \pt\Delta\pt$ are similar between each other and thereby essentially cancel one another  leading to 
a suppression of the apparent strength of particle correlations in that
range. At small values of $\Delta\eta$ and $\Delta\varphi$, pairs with
$\Delta \pt\Delta \pt>0$ clearly dominate and yield the narrow peak
observed. This behavior is consistent with the angular ordering expected in
jets as confirmed by model calculations based on
PYTHIA8~\cite{Sahoo:2018uhb}. The effect is, however, also expected
from hadronic resonance decays. It should be additionally noted that a similar pattern was observed for $R_2^{\rm CI}$ and $P_2^{\rm CI}$ correlation functions 
measured in p--Pb and Pb--Pb collisions~\cite{AliceDptDptLongPaper}.

\subsection{Charge-Dependent (CD) correlation functions}
\label{sec:CD}
 
\begin{figure}[tb]
  \includegraphics[scale=0.37]{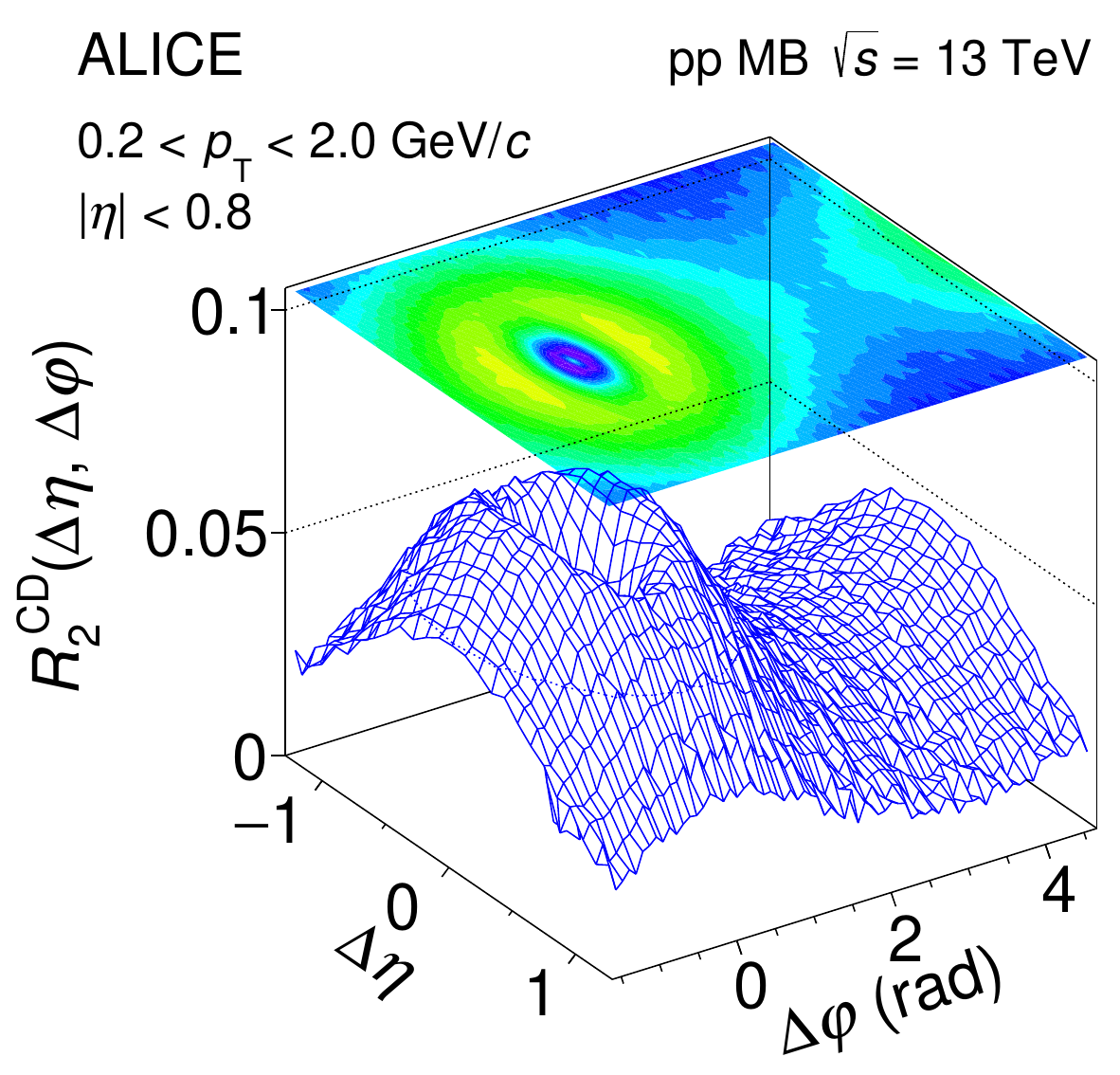}
   \qquad
  \includegraphics[scale=0.37]{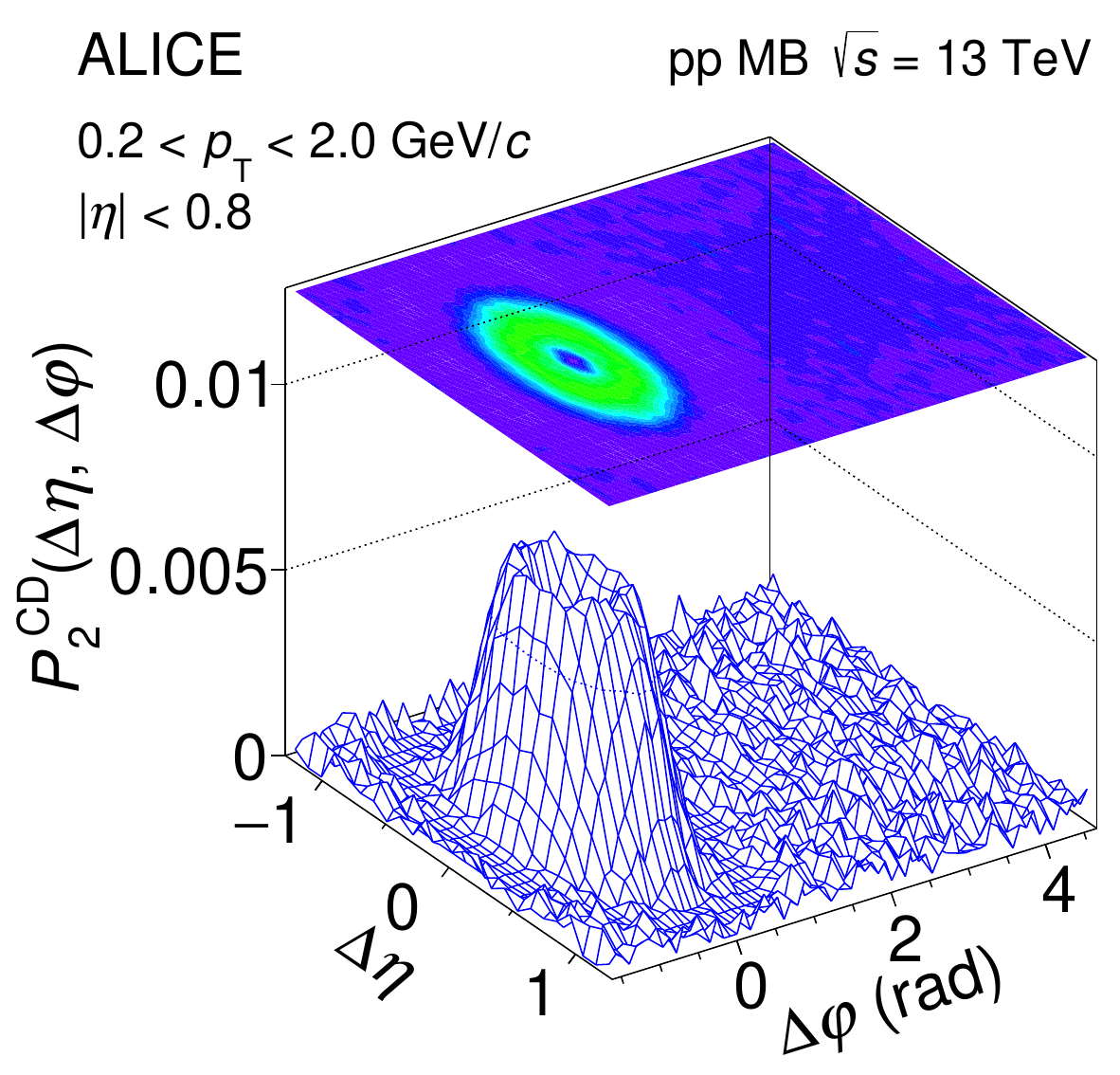}   
  \caption{Correlation functions $\RtwoCD$ (left ) and $\PtwoCD$
    (right) of charged hadrons  in minimum bias pp collisions at $\s$ = 13 TeV. Charged hadrons are selected in the  range $0.2 < \pt < 2.0$ GeV/$c$ and with pseudorapidity $|\eta| < 0.8$.}%
  \label{fig:r2p2CdSys}%
\end{figure}

\begin{figure}[tb]
  \centering
  \includegraphics[scale=0.7]{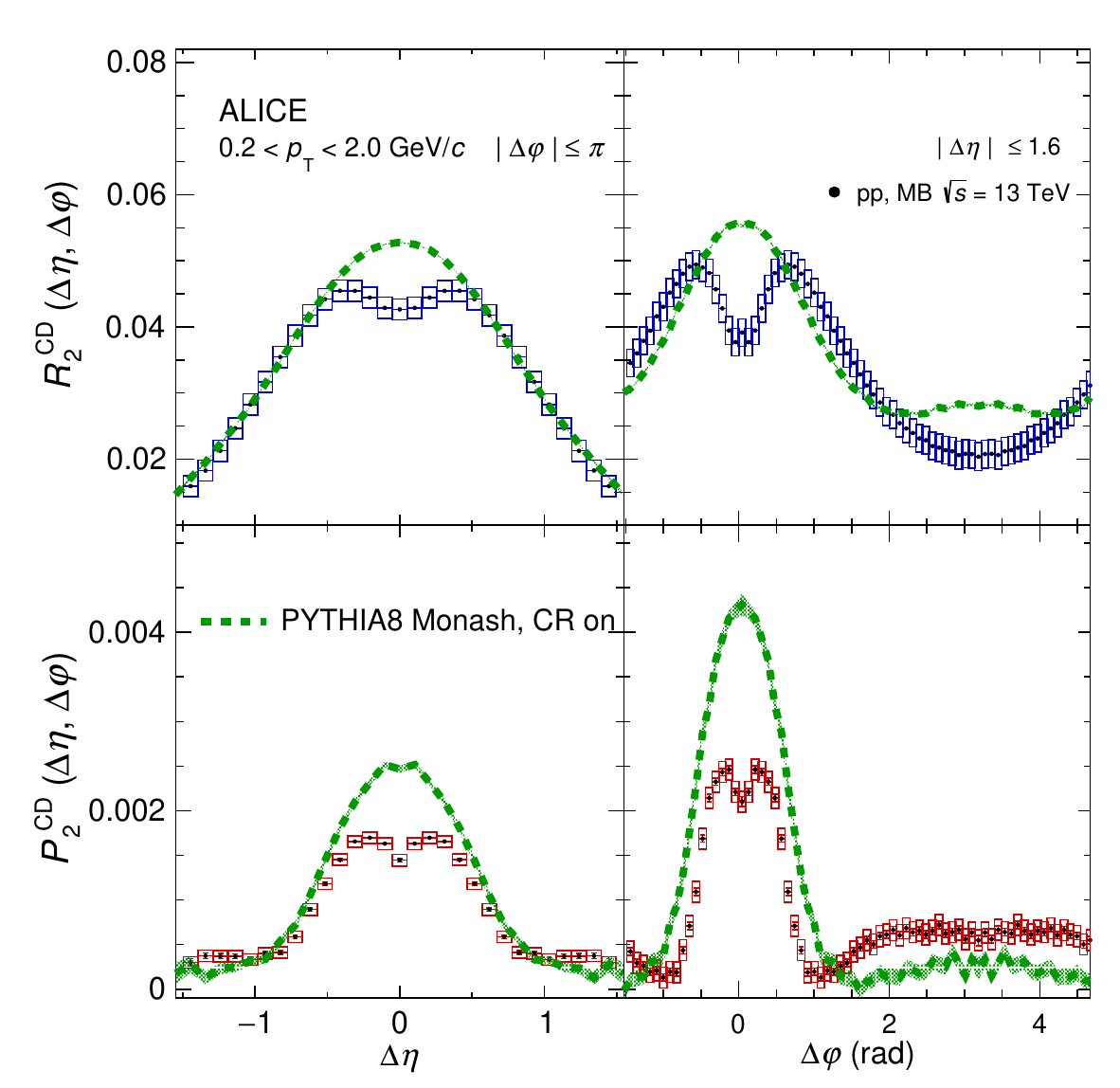}   
   \caption{Projections onto the $\Deta$ (left) and $\Dphi$ (right) axes of the $\RtwoCD$ (top)
  and $\PtwoCD$ (bottom) correlation functions shown in Fig.~\ref{fig:r2p2CdSys}. The $\Deta$ and $\Dphi$ projections are calculated as averages of the two-dimensional 
  correlations in the range $|\Dphi| \leq \pi$ and $|\Deta| \leq 1.6$, respectively. Vertical bars and
boxes represent statistical and systematic uncertainties,
respectively. Simulations using PYTHIA8 with the Monash 2013 tune and color reconnection (CR) enabled, as described in the text, are shown as green dashed lines.}
 \label{fig:r2p2CdProjData}
\end{figure}

The $\RtwoCD$ and $\PtwoCD$ correlation functions are shown as two-dimensional plots in 
the left   and  right panels of  Fig.~\ref{fig:r2p2CdSys}, respectively, and their projections 
onto $\Deta$ and $\Dphi$ axes are displayed in Fig.~\ref{fig:r2p2CdProjData}. It can be noted that the near-side peak of both correlation functions feature  a narrow dip around $(\Deta, \Dphi) = (0,0)$, 
which stems in large part due to Bose--Einstein Quantum Statistics. In addition, the near-side peak of the $P_2^{\rm CD}$ correlation function is substantially narrower than that of $\RtwoCD$. This difference is expected from the angular and transverse momentum ordering effect already mentioned~\cite{Sahoo:2018uhb}.

At LHC energies, positively and negatively charged particles are produced in nearly equal multiplicities and feature similar $\pt$ spectra ~\cite{partProd1}. In large  collision systems, negatively and positively charged particles are found to have nearly equal momentum anisotropy distributions. It entails that in large systems, the $\RtwoCD$ and $\PtwoCD$ correlation functions feature an essentially flat and featureless away-side across a wide interval of collision centrality spanning from the most central collisions down to peripheral collisions. While particle momentum anisotropies, possibly originating from a collectively expanding system, have been observed in high-multiplicity pp collisions, it is expected that non-flow correlations, such as those resulting from energy--momentum conservation (causing back-to-back particle emission in the transverse plane), should impact LS and US charged particle pairs similarly and thus lead to a flat away-side distribution. It can be observed, however, that the away-side of the $\RtwoCD$ correlation function is not flat and shows a dependence on both $\Deta$ and $\Dphi$ which reflects a small difference in the away-side emission of LS and US pairs. This likely results from the rather broad nature of the near-side peak, which extends beyond $\Dphi$ = $\pi/2$ towards the away-side. In contrast, however, the $\PtwoCD$ near-side peak is narrower and does not extend on the away-side. This correlation function thus has a rather flat away-side. This behavior was also observed in p--Pb and Pb--Pb collisions~\cite{AliceDptDptLongPaper}.

\subsection{\texorpdfstring{$R_2$}{r2}
and \texorpdfstring{$P_2$}{p2} correlation functions with PYTHIA8 }
The PYTHIA8~\cite{Skands:2014pea} event generator has had great successes in modeling recent measurements of single particle and jet production in high-energy pp collisions. It is thus of interest to investigate whether it can also reproduce the measured $\Rtwo$ and $\Ptwo$ correlation functions. Comparisons with PYTHIA8 calculations of CI and CD components of these correlation functions are presented in Figs.~\ref{fig:r2p2CiProjData} and~\ref{fig:r2p2CdProjData}, respectively. Focusing on the $\Delta\eta$ (left) and $\Delta\varphi$ (right) projections of CI correlation functions, shown in Fig.~\ref{fig:r2p2CiProjData}, it can be noted  that PYTHIA8 essentially captures the basic features of the $\RtwoCI$ and $\PtwoCI$ correlation functions. The projections indeed show near-side peaks and away-side structures that approximately match those observed experimentally. The small differences between the calculations and the measurements may originate from various phenomena, the detailed analysis of which is beyond the scope of this work. 

Switching the discussion to the projections of the CD correlations, shown in Fig.~\ref{fig:r2p2CdProjData}, it can be observed that also in this case, PYTHIA8 captures the salient features of the measured correlation functions, with  one notable exception: the dip structures found at $\Delta\eta=0$ and $\Delta\varphi=0$ in experimental data. By construction, CD correlations are sensitive to the presence of HBT, manifested in LS, and  known to exist in  all colliding systems. The HBT correlations, typically measured in terms of invariant momentum differences, are expected to be relevant at small $\Delta\eta$, $\Delta\varphi$ separation~\cite{ALICE:2015hav}. 
As such the width of the dip in the CD correlations reflects in part  the width of the peak in LS correlations due to Bose--Einstein condensation and thus serve as an indicator of the system size. Such effects are not seen in the PYTHIA8 predictions shown in the figures because no HBT afterburner was used in the calculations~\cite{Pratt:2018ebf}. 
It is also possible that the dip observed at small pair separation may arise from other causes related to particle production from hadronic resonance decays or other sources of charge conserving production mechanisms. Furthermore, PYTHIA8 also predicts the presence of a small bump structure centered at $\Delta\varphi=\pi$ in $\Delta\varphi$ projections of $\RtwoCD$ which is not observed in the data. Such small away-side peak is likely caused by momentum conservation effects which manifest themselves by back-to-back particle emission at low multiplicity. Similarly,  PYTHIA8 reproduces the $\PtwoCD(\Deta)$ with some deviations near $\Deta = 0$ and towards the edge. Moreover, it can be observed that in contrast to the $\RtwoCD$, where PYTHIA8  approximately matches the overall magnitude of the correlation function, the near-side peak of the $\PtwoCD$ is overestimated by a factor of two while the away-side magnitude is underestimated by the same factor. These results show that while PYTHIA8 qualitatively reproduces the measured correlation functions. 

\subsection{Evolution of the near-side peak width with multiplicity in different collision systems}
\label{sec:width}
\begin{figure}[!htb]
  \centering
  \includegraphics[scale=0.65]{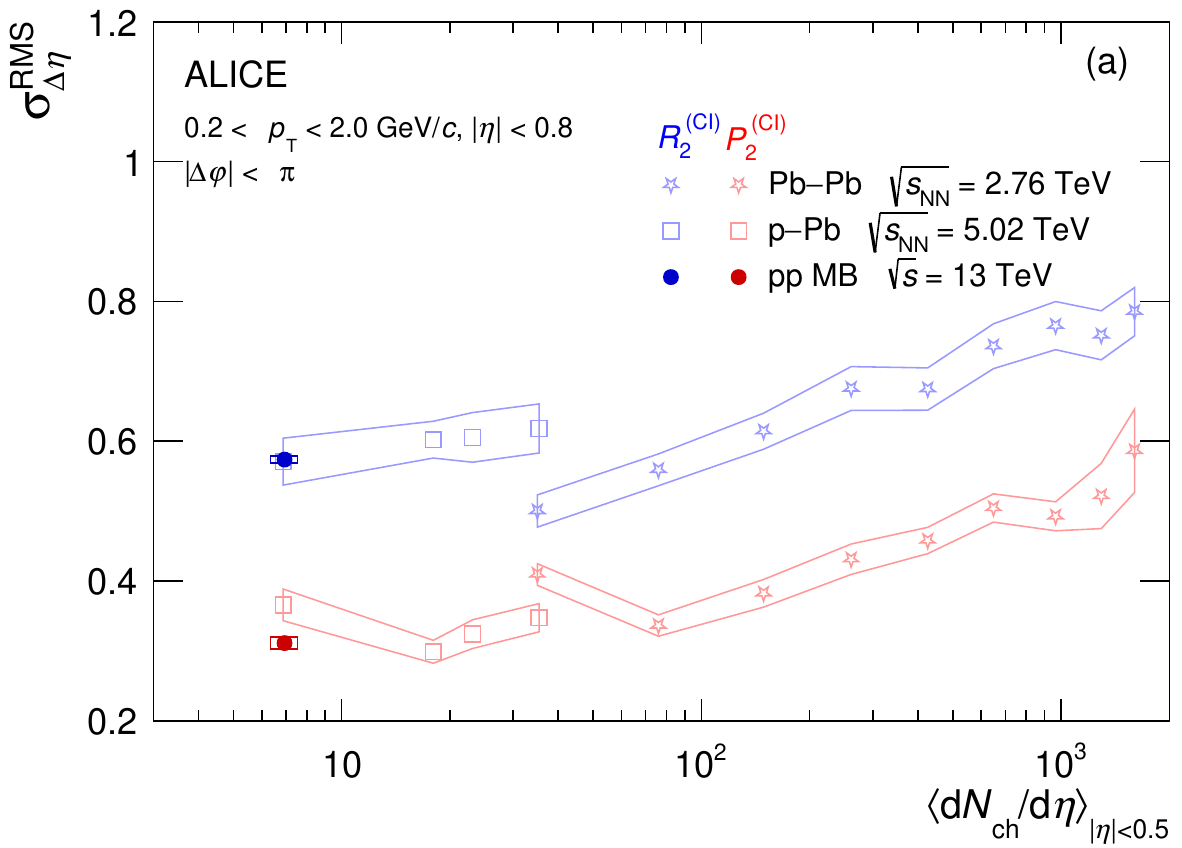}   
   \includegraphics[scale=0.65]{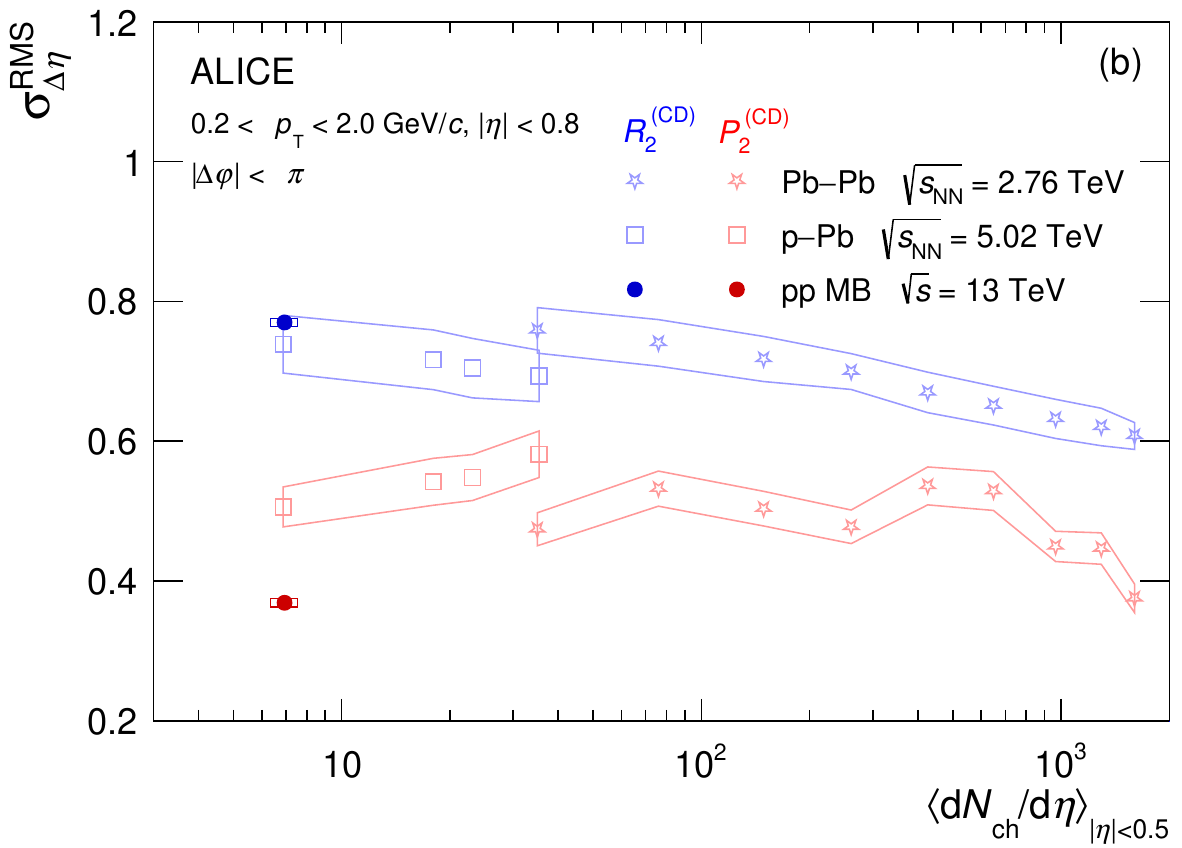}
 \caption{Width of (a) $\RtwoCI(\Deta)$ (blue markers) and $\PtwoCI(\Deta)$ (red markers) and (b) $\RtwoCD(\Deta)$ (blue markers) and $\PtwoCD(\Deta)$ (red markers) correlation functions along $\Deta$ measured within $|\Dphi| < \pi$  in pp collisions and compared with results from p--Pb and Pb--Pb collisions~\cite{AliceDptDptLongPaper} as a function of $\dnavg$. Vertical bars and boxes represent statistical and systematic uncertainties, respectively.}
  \label{fig:widthCiDeta}
\end{figure}

\begin{figure}[tb]
  \centering
  \includegraphics[scale=0.65]{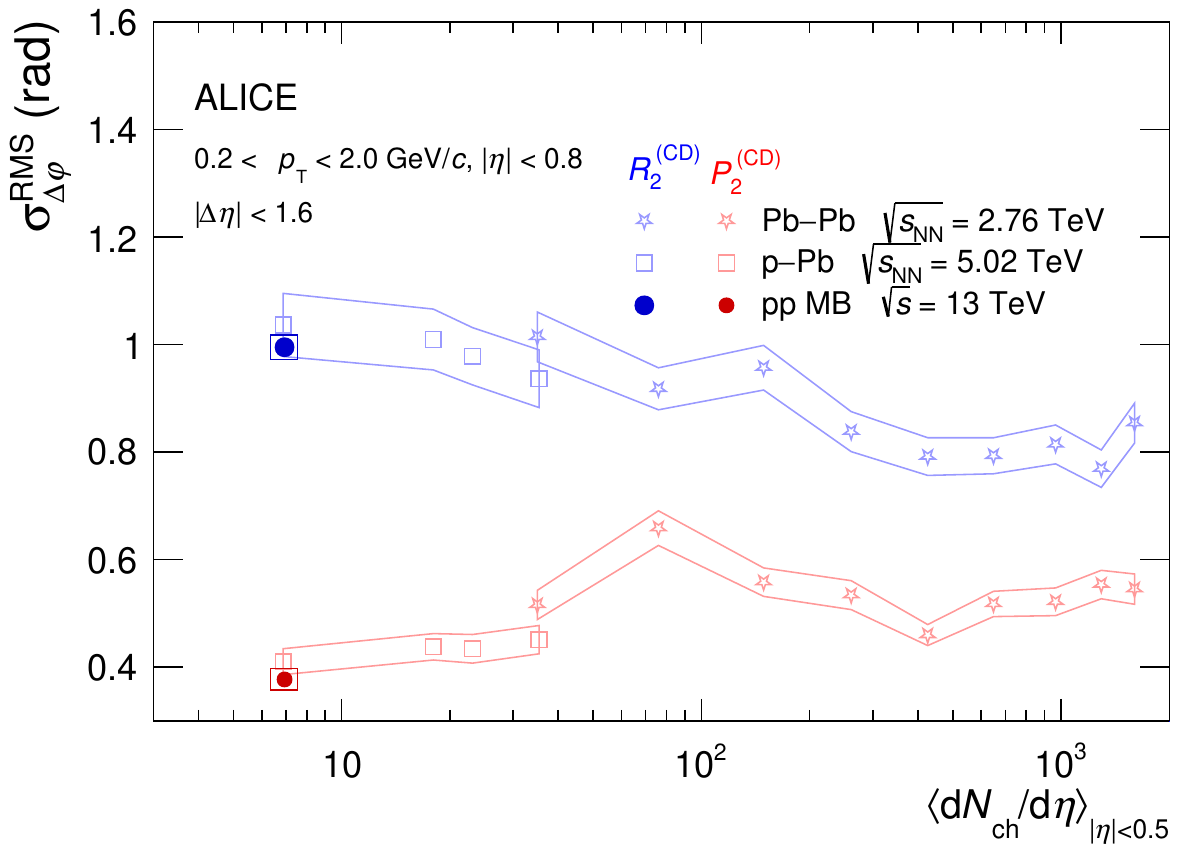}
  
  \caption{Widths of $\RtwoCD(\Dphi)$ (blue markers) and $\PtwoCD(\Dphi)$
    (red markers) correlation functions along $\Dphi$
measured within $|\Deta| < 1.6$  for pp and within $|\Deta| < 1.8$
for p--Pb and Pb--Pb collisions~\cite{AliceDptDptLongPaper} as a function of
$\dnavg$. Vertical bars and boxes represent statistical and systematic uncertainties, respectively.  }
  \label{fig:widthDphi}
\end{figure}

The widths of the near-side peak of the measured correlation functions  along the $\Delta\eta$ and $\Delta\varphi$ axes are commonly used to characterize  the many processes that contribute to the  strength and shape of particle correlations. It is interesting, in particular, to consider how these widths evolve with the produced particle multiplicity quantified by the charged-particle pseudorapidity density at midrapidity ($\dnavg$), the collision system size, as well as the collision energy. 

This section reports the near-side peak root mean square (RMS) widths of correlation functions computed using Eq.~(\ref{eq:widths}) and  shown in Figs.~\ref{fig:widthCiDeta} and~\ref{fig:widthDphi}. Given some  
ambiguities existing in the identification of a baseline for broad peaks measured in a narrow acceptance, as in the case of the $\Delta\eta$ projections of $R_2^{\rm CI}$ and $P_2^{\rm CI}$, the RMS widths are computed  using  two methods: the first involves accounting for the presence of a baseline (or correlation plateau) underneath the peak and the second characterizes the width of the peak in a fixed $\Delta\eta$ or $\Delta\varphi$ range. In the case of the first method, since a plateau is not evident based on the shape of the projected correlation functions (reported in Fig.~\ref{fig:r2p2CiProjData}), the correlation strength at the edges of the acceptance is used as baseline and thus constitutes a somewhat arbitrary definition of the baseline, which ignores the correlation strength at $\Delta\eta$ values beyond the experimental acceptance. It nonetheless provides an estimator of the width of interest albeit with a bias. The second method has merits and limitations as well. It evidently does not suffer from the somewhat arbitrary requirement of a baseline determination, but its magnitude is largely defined by the experimental acceptance itself. For instance, the offset for $\PtwoCI (\Dphi)$, illustrated in Fig.~\ref{fig:r2p2CiProjData}, is determined by averaging three consecutive bins at the turning points near the edges of the near-side peak. Indeed, for a shallow peak or no peak whatsoever, the RMS width would be equal to the accepted range divided by $\sqrt{12}$. 

The RMS widths of the $R_2$ and $P_2$ correlation functions computed   with and without offsets (offset = 0) are reported in Tab.~\ref{tab:widthOffset}. Despite the fact that both methods mentioned above use an offset, the table enables a direct comparison between calculations performed with and without this offset. To maintain consistency with previous results in \pPb~and~\PbPb,~ the widths for $\RtwoCD$ 
  are considered without applying any offset. The computed RMS widths  feature a strong dependence  on the method chosen for their evaluation. Comparisons with the results of prior measurements and/or with model predictions must then be carried out in a consistent manner. 
Figures~\ref{fig:widthCiDeta} and ~\ref{fig:widthDphi} compare the widths along $\Deta$ and $\Dphi$ obtained in pp collisions at $\s$ = 13 TeV, with those reported by the ALICE Collaboration in  p--Pb collisions at $\snn$ = 5.02
TeV, and Pb--Pb collisions at $\snn$ = 2.76 TeV~\cite{AliceDptDptLongPaper}. In the case of $\RtwoCD (\Deta)$, the RMS widths computed without offset are shown in Fig.~\ref{fig:widthCiDeta}, considering the fact that the correlation vanishes at large $|\Deta|$ values, whereas finite offsets were used for other  projections, consistent with the previous results in p--Pb and Pb--Pb collisions~\cite{AliceDptDptLongPaper}. 

\begin{table}[tb]
  \centering
    \caption{Comparison of the widths calculated with  and without the offset (T = 0) in Eq.~\ref{eq:widths}.}
\begin{tabular}{cccc}
    \hline
\hline    
    Correlator
    & Category
    & \Centerstack{ With Offset} & \Centerstack{ Without Offset}\\
    \hline
$\RtwoCI$
& \Centerstack{ $\Deta$ \\ $\Dphi$ }
    & \Centerstack{ $0.57\pm 0.0003 (\text{stat.})\pm 0.0052 (\text{syst.})$\\ $0.54\pm 0.0002 (\text{stat.})\pm 0.0086 (\text{syst.})$}
    & \Centerstack{ $0.90\pm < 0.0001 (\text{stat.})\pm 0.0079 (\text{syst.})$\\ $0.93\pm < 0.0001 (\text{stat.})\pm 0.0148 (\text{syst.})$}\\

$\PtwoCI$
& \Centerstack{ $\Deta$ \\ $\Dphi$ }
    & \Centerstack{ $0.31\pm 0.0007 (\text{stat.})\pm 0.041 (\text{syst.})$\\ $0.33\pm 0.0007 (\text{stat.})\pm 0.004 (\text{syst.})$}
    & \Centerstack{ $0.51\pm 0.0001 (\text{stat.})\pm 0.079 (\text{syst.})$\\ $0.51\pm 0.0002 (\text{stat.})\pm 0.006 (\text{syst.})$}\\

$\RtwoCD$
  & \Centerstack{ $\Deta$ \\ $\Dphi$ }
    & \Centerstack{ $0.65\pm 0.001 (\text{stat.})\pm 0.0058 (\text{syst.})$\\ $0.99\pm 0.0025 (\text{stat.})\pm 0.023 (\text{syst.})$}
    & \Centerstack{ $0.77\pm 0.0004 (\text{stat.})\pm 0.0069 (\text{syst.})$\\ $1.27\pm 0.001 (\text{stat.})\pm 0.0298 (\text{syst.})$}\\

$\PtwoCD$
& \Centerstack{ $\Deta$ \\ $\Dphi$ }
    & \Centerstack{ $0.37\pm 0.0039 (\text{stat.})\pm 0.0062 (\text{syst.})$\\ $0.37\pm 0.0031 (\text{stat.})\pm 0.02 (\text{syst.})$}
    & \Centerstack{ $0.51\pm 0.0024 (\text{stat.})\pm 0.0084 (\text{syst.})$\\ $0.41\pm 0.0025 (\text{stat.})\pm 0.0215 (\text{syst.})$}\\
\hline
 \hline
  \end{tabular}    \label{tab:widthOffset}   
\end{table}

Figure~\ref{fig:widthCiDeta} contrasts  the longitudinal near-side  widths  of  CI and CD correlation functions measured in pp collisions with those observed in p--Pb, and Pb--Pb collisions~\cite{AliceDptDptLongPaper} as a function of  the density of charged particles  
in the transverse momentum interval 0.15 $ < \pt < $ 20 $\gevc$ and within the pseudorapidity range, $|\eta|<0.5$~\cite{rho1}. 
In the top panel of Fig.~\ref{fig:widthCiDeta}, it can be observed that the longitudinal RMS  widths of 
$\RtwoCI(\Deta)$ and $\PtwoCI(\Deta)$ tend to rise  with increasing particle density in p--Pb and Pb--Pb collisions, a feature that may be indicative of the finite shear viscosity of the medium created in these collisions, especially  in Pb--Pb collisions~\cite{G2PbPb,Gonzalez:2020bqm}. The longitudinal widths of $R_2^{\rm CD}$ measured in pp collisions are in a good agreement with those obtained in low multiplicity p--Pb collisions. In sharp contrast to that is a large deviation of the width of $P_2^{\rm CD}$ near-side peak from the value measured in p--Pb collisions with similar charged particle multiplicity. While the origin of this discrepancy is not entirely clear, it may in part arise from the difference in collision energy as well as mean-$\pt$~\cite{ALICE:2013rdo} between pp and p--Pb collisions. 
 
The evolution of the near-side peak width of
$\RtwoCD(\Dphi)$ correlation functions with produced particle multiplicity measured  in pp, p--Pb, and Pb--Pb collisions is shown in Fig.~\ref{fig:widthDphi}. The observed decrease of the width  from low to high multiplicity is likely due to the rise of the (transverse) radial flow in these collisions although it is also expected that the diffusivity of the matter produced in Pb--Pb collisions might induce a small broadening of the correlation function~\cite{Pratt:2015jsa,Pratt:2019pnd}. The multiplicity dependence of  the $\PtwoCD$ correlation function   exhibits a  more complicated behavior. The azimuthal width of $P_2^{\rm CD}$ displays a small rise from the multiplicity observed in  minimum bias pp collisions to the largest multiplicities observed in p--Pb collisions. This rise continues up to $\dnavg \approx  100$ in Pb--Pb, but is followed by a modest decrease at higher multiplicities. Again, it can be noted that  the near-side peak of $\PtwoCD$ is considerably narrower than the one of $\RtwoCD$.  Overall, it is visible that the azimuthal width of the $P_2^{\rm CD}$ near-side peak is smaller in pp and p--Pb than in Pb--Pb collisions. This suggests that the impact of the $p_{\rm T}$ angular ordering effect is weaker at high multiplicity perhaps owing to a greater probability of multiple scattering in such larger systems. 

The evolution of the near-side peak widths of the $R_2^{\rm CI,CD}$,  $P_2^{\rm CI,CD}$ correlators from pp, to p--A, to A--A collisions is somewhat puzzling. This may be in part because the strength of both the $R_2$ and $P_2$ correlation functions is manifestly non vanishing outside the experimental acceptance but  cannot be reliably extrapolated thereby leading to ambiguities in the use of ``offsets" employed in the computation of the widths. The apparent mismatch might also arise from differences in the role on femtoscopic correlations in these three systems.

\subsection{Balance function and its integral}
\label{sec:bf}   
Charge conservation dictates that the production of a positively charged particle must be accompanied by the creation of a charge balancing particle, i.e., a negatively charged particle. However, both    production and transport mechanisms of charged particle pairs may impact the shape and strength of  the  $R_2^{\rm CD}$  correlation function. A focus on the particle production itself can be brought about by considering a measurement of  charge balance function ($B$), which in the context of collisions at the LHC energies can be written: $B = \rho^{\mathrm{B}} \RtwoCD$, where $\rho^{\mathrm{B}}$ represents the charged particle density calculated in the kinematic range of $|\eta| <$ 0.8 and 0.2 $< \pt <$ 2 $\gevc$.  This single particle density is obtained by integration of  prior ALICE measurements of the differential invariant yield of charged particles~\cite{rho1}
 according to 
\be
\label{eq:bfFact}
\rho^{\mathrm{B}} = \Big\langle\frac{\mathrm{d^{2}}N_\mathrm{ch}}{\mathrm{d}\varphi\mathrm{d}\eta} \Big\rangle = \int_{p_{\mathrm {T,min}}}^{p_{\mathrm {T,max}}} \frac{\mathrm{d^{2}}N_\mathrm{ch}}{2\pi p_{\mathrm T} \mathrm{d} p_{\mathrm T} \mathrm{d}\eta} \times p_{\mathrm T} \mathrm{d} p_{\mathrm T}.
\ee

\begin{figure}[tb]
    \centering
    \includegraphics[width=8cm,height=16cm,keepaspectratio]{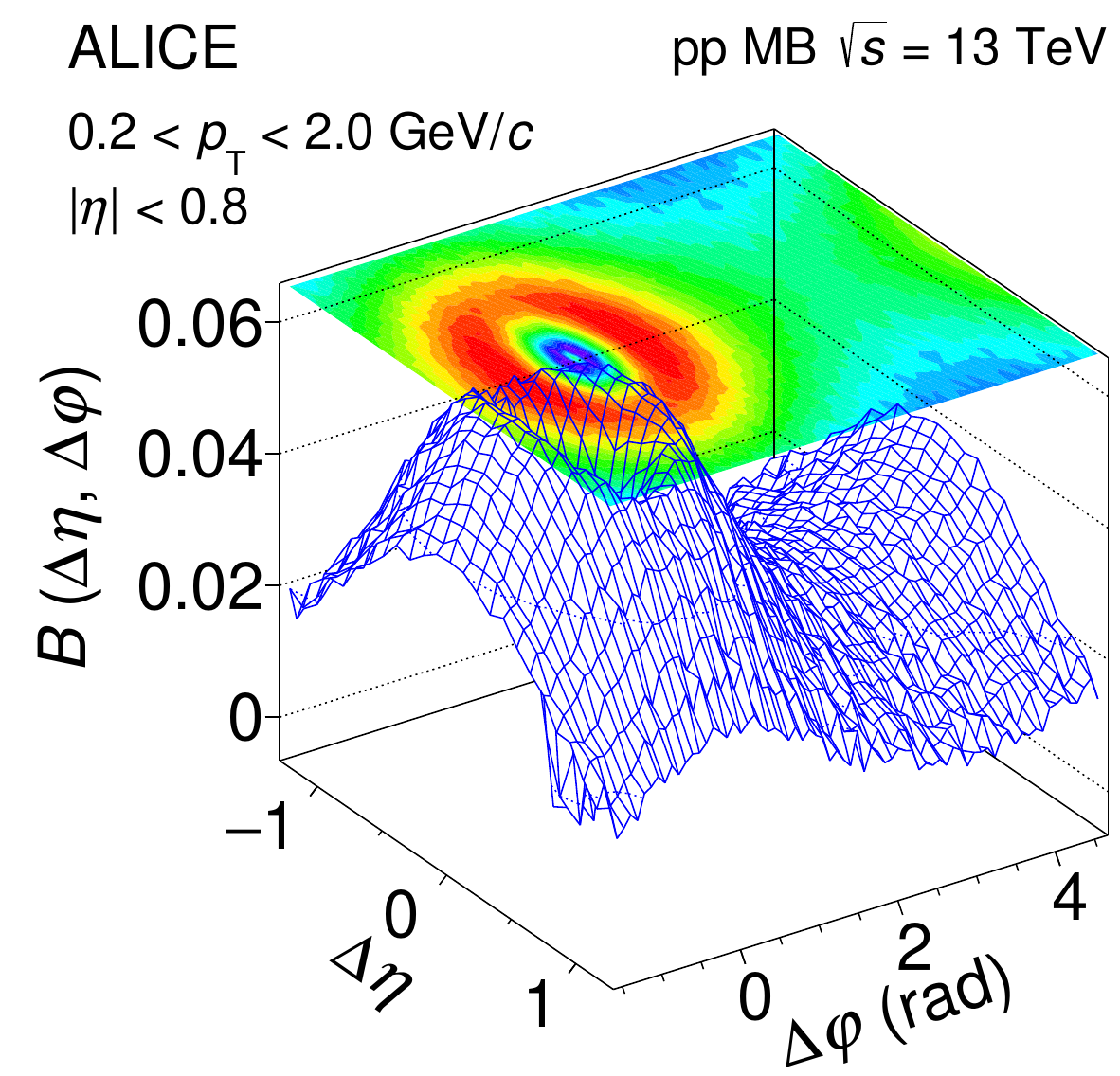}
    \hspace{0.5cm}
    \includegraphics[width=7cm,height=14cm,keepaspectratio]{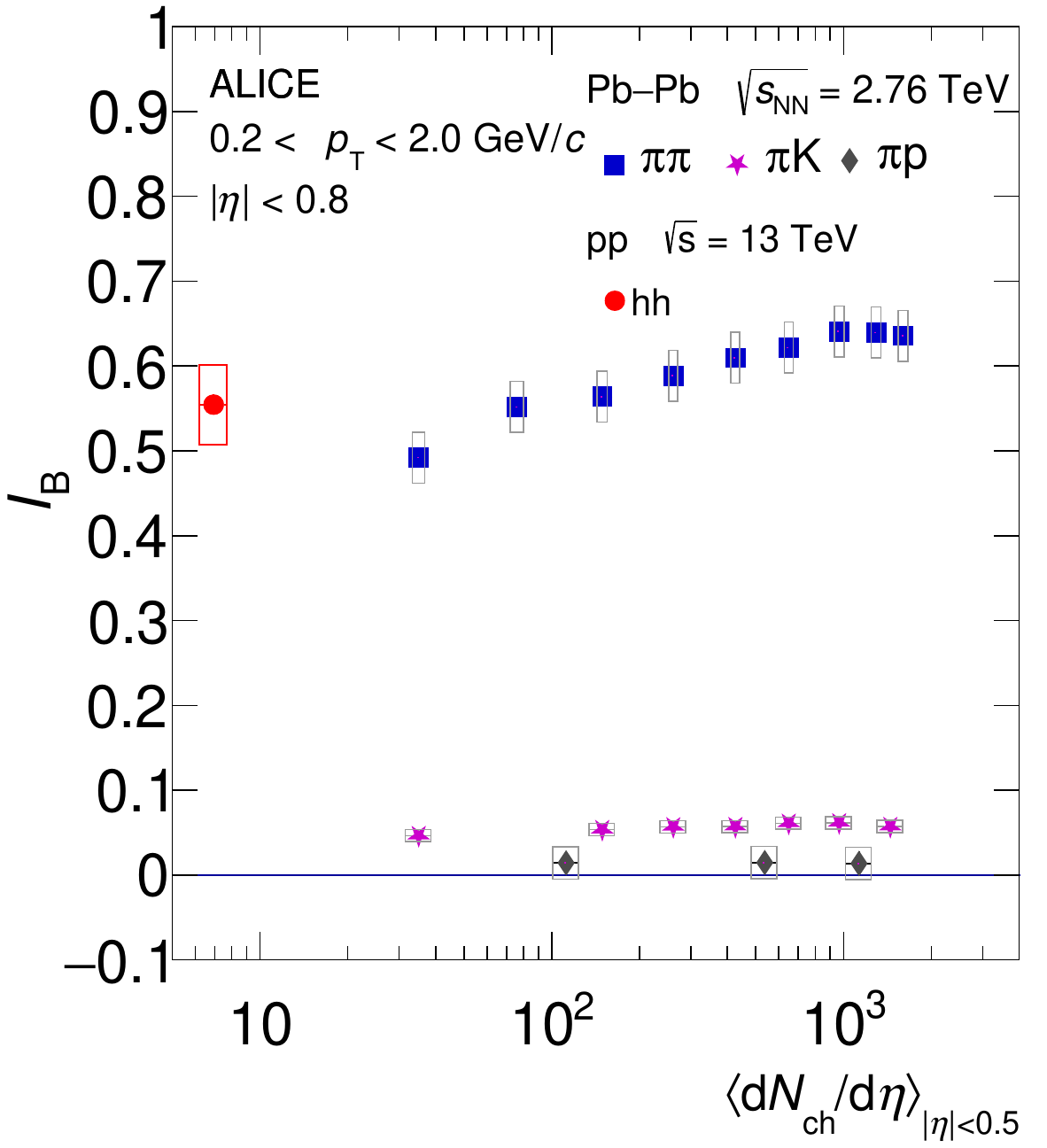}   

  \caption{Balance function (left panel) of charged hadrons, in the selected $\pt$
    range, obtained using ALICE data in pp collisions at $\s$ = 13
    TeV. Integral of $B$ (right panel) results for charged hadrons in pp
    collisions at $\s$ = 13 TeV compared with the $B$ integral of identified hadron ($\pi \pi$, $\pi$K, and $\pi$p) pairs in Pb--Pb collisions at $\snn$ = 2.76 TeV~\cite{genBF}. Vertical bars and boxes represent statistical and systematic uncertainties, respectively.}%
  \label{fig:bf}%
\end{figure}
The balance function, plotted  as a function of $\Deta$ and $\Dphi$ in the left panel of Fig.~\ref{fig:bf}, has the same shape of the $\RtwoCD$ correlation function reported in Fig.~\ref{fig:r2p2CdSys}(a) modulo a rescaling factor. The integral of the charge balance function, computed according to Eq.~\ref{eq:IB}, amounts to $0.55\pm 0.0039 (\text{stat.})\pm 0.06 (\text{syst.})$ and it is shown in the right panel of Fig.~\ref{fig:bf}. This means that given a positive hadron is measured in the acceptance $|\eta| < 0.8$, $0.2 < p_{\mathrm T}\ < 2.0$ GeV/$c$,  the chance of 
finding a charge balancing negative hadron in that same particle acceptance amounts to  55$\%$. The probability is evidently smaller than unity because the charge  balancing hadron may be emitted (``leak") outside the acceptance, i.e., either at pseudorapidity $|\eta|>0.8$ or outside the $0.2 < p_{\rm T} < 2.0$ GeV/$c$ range of the measurement. It is worth noting, however,  that this probability 
is compatible to the sum of the probabilities (in  essentially identical acceptances)  of observing $\pi^{-}$, K$^{-}$, and $\overline{\rm{p}}$ as balancing partners of a $\pi^{+}$ in Pb--Pb collisions at $\snn = 2.76$ TeV~\cite{genBF}. In this context, $\pi\pi$ pairs are predominantly observed, while contributions from other pairs are negligible. Nevertheless, the transverse momentum ranges for identified particles are different, and may account for the small modification. More detailed measurements of balance functions of identified hadrons in small and large collision systems and their evolution with produced particle multiplicity are forthcoming. One should also remark that the integral of the balance function measured in this work deviates appreciably from balance function integrals reported by the CMS collaboration based on their recent measurements~\cite{CMS:2023sua}. While the CMS measurements benefit from a much larger rapidity acceptance, $|\eta_{\rm cms}|<2.4$, they do not feature the low $p_{\rm T} = 0.2$ GeV/$c$ achieved within the ALICE measurements reported in this work. The CMS measurements also use a momentum ordering $p_{\rm T,assoc} < p_{\rm T,trig}$ that may impact the magnitude of the integral of balance functions given a charge balancing partner may also be  found at $p_{\rm T,assoc} \ge  p_{\rm T,trig}$. 

\section{Summary}
\label{sec:summary}

Two-particle differential number correlation function, $\Rtwo\DetaDphi$, and  transverse
momentum correlation function, $\Ptwo\DetaDphi$, of charged particles in the transverse momentum range 0.2 $< \pt <$ 2.0 $\gevc$ and pseudorapidity range $|\eta| <$ 0.8 are measured in minimum-bias pp collisions at a center-of-mass energy of $\s$ = 13 TeV. Correlation functions are studied individually for LS and US particle pairs  and combined to obtain CI and
CD. These correlation functions exhibit features similar to those observed in larger collision systems, including, in particular, a somewhat narrow,  but prominent peak at small particle pair separation $\Delta\eta$, $\Delta\varphi$. As well as in prior measurements in p--Pb and Pb--Pb collisions, the near-side peak of both CI and CD $P_2$ correlations is found to be significantly narrower than those of $R_2$ correlations, a feature expected from momentum versus angular ordering in jets and hadronic resonance decays. The $R_2^{\rm CD}$ and $P_2^{\rm CD}$  exhibit a narrow dip at the center of their respective near-side peaks. This feature is understood to result in part from Bose--Einstein correlations of LS particles and the relatively small source size of pp collisions. The projections of measured $\Rtwo\DetaDphi$
and $\Ptwo\DetaDphi$ onto $\Deta$ and $\Dphi$ are compared
with PYTHIA8 calculations. It is found that PYTHIA8 qualitatively reproduces measurements in pp collisions, but cannot catch all the observed features suggesting that further improvement of the model is needed.

The RMS widths of the near-side peak of $R_2^{\rm CI}$ along the longitudinal and $R_2^{\rm CD}$ along the longitudinal and azimuthal  directions are found to be in good agreement with the widths of correlation functions observed at low charged particle multiplicity in p--Pb collisions. This agreement indicates the breadth of these correlation functions likely  primarily depends on the produced multiplicity. However, the widths of $P_2^{\rm CI}$ and
$P_2^{\rm CD}$ in the longitudinal direction observed in pp  are significantly smaller than those measured in p--Pb collisions. The azimuthal width of $P_2^{\rm CD}$ observed in pp approximately matches the width measured in p--Pb at similar charged particle density. The widths of the near-side peaks of $\Rtwo$ and $\Ptwo$ correlation functions  are studied in detail and compared with the measurements performed in larger colliding systems.   The widths of the near-side peak of 
$\Rtwo$ for CI and CD combinations follow a consistent trend despite of
differences in the center-of-mass energy for the three systems. However, the widths of the near-side peak of $\Ptwo (\Deta)$ for CI and CD show some deviations in the  three systems. These deviations may in part come from differences in collision energy~\cite{ALICE:2013rdo}.

Furthermore, the integral of the balance function ($I_{\mathrm{B}}$) is measured in pp collisions based on the $R_2^{\rm CD}$ correlation function. The magnitude of the integral is  found to be compatible to the sum of balance function integrals measured for individual charged hadron pairs, $\pi \pi$, $\pi$K, and $\pi$p, in Pb--Pb collisions by the ALICE Collaboration. This suggests that balance functions and their integrals evolve rather slowly with system size and collision energy.


\newenvironment{acknowledgement}{\relax}{\relax}
\begin{acknowledgement}
\section*{Acknowledgements}

The ALICE Collaboration would like to thank all its engineers and technicians for their invaluable contributions to the construction of the experiment and the CERN accelerator teams for the outstanding performance of the LHC complex.
The ALICE Collaboration gratefully acknowledges the resources and support provided by all Grid centres and the Worldwide LHC Computing Grid (WLCG) collaboration.
The ALICE Collaboration acknowledges the following funding agencies for their support in building and running the ALICE detector:
A. I. Alikhanyan National Science Laboratory (Yerevan Physics Institute) Foundation (ANSL), State Committee of Science and World Federation of Scientists (WFS), Armenia;
Austrian Academy of Sciences, Austrian Science Fund (FWF): [M 2467-N36] and Nationalstiftung f\"{u}r Forschung, Technologie und Entwicklung, Austria;
Ministry of Communications and High Technologies, National Nuclear Research Center, Azerbaijan;
Conselho Nacional de Desenvolvimento Cient\'{\i}fico e Tecnol\'{o}gico (CNPq), Financiadora de Estudos e Projetos (Finep), Funda\c{c}\~{a}o de Amparo \`{a} Pesquisa do Estado de S\~{a}o Paulo (FAPESP) and Universidade Federal do Rio Grande do Sul (UFRGS), Brazil;
Bulgarian Ministry of Education and Science, within the National Roadmap for Research Infrastructures 2020-2027 (object CERN), Bulgaria;
Ministry of Education of China (MOEC) , Ministry of Science \& Technology of China (MSTC) and National Natural Science Foundation of China (NSFC), China;
Ministry of Science and Education and Croatian Science Foundation, Croatia;
Centro de Aplicaciones Tecnol\'{o}gicas y Desarrollo Nuclear (CEADEN), Cubaenerg\'{\i}a, Cuba;
Ministry of Education, Youth and Sports of the Czech Republic, Czech Republic;
The Danish Council for Independent Research | Natural Sciences, the VILLUM FONDEN and Danish National Research Foundation (DNRF), Denmark;
Helsinki Institute of Physics (HIP), Finland;
Commissariat \`{a} l'Energie Atomique (CEA) and Institut National de Physique Nucl\'{e}aire et de Physique des Particules (IN2P3) and Centre National de la Recherche Scientifique (CNRS), France;
Bundesministerium f\"{u}r Bildung und Forschung (BMBF) and GSI Helmholtzzentrum f\"{u}r Schwerionenforschung GmbH, Germany;
General Secretariat for Research and Technology, Ministry of Education, Research and Religions, Greece;
National Research, Development and Innovation Office, Hungary;
Department of Atomic Energy Government of India (DAE), Department of Science and Technology, Government of India (DST), University Grants Commission, Government of India (UGC) and Council of Scientific and Industrial Research (CSIR), India;
National Research and Innovation Agency - BRIN, Indonesia;
Istituto Nazionale di Fisica Nucleare (INFN), Italy;
Japanese Ministry of Education, Culture, Sports, Science and Technology (MEXT) and Japan Society for the Promotion of Science (JSPS) KAKENHI, Japan;
Consejo Nacional de Ciencia (CONACYT) y Tecnolog\'{i}a, through Fondo de Cooperaci\'{o}n Internacional en Ciencia y Tecnolog\'{i}a (FONCICYT) and Direcci\'{o}n General de Asuntos del Personal Academico (DGAPA), Mexico;
Nederlandse Organisatie voor Wetenschappelijk Onderzoek (NWO), Netherlands;
The Research Council of Norway, Norway;
Pontificia Universidad Cat\'{o}lica del Per\'{u}, Peru;
Ministry of Science and Higher Education, National Science Centre and WUT ID-UB, Poland;
Korea Institute of Science and Technology Information and National Research Foundation of Korea (NRF), Republic of Korea;
Ministry of Education and Scientific Research, Institute of Atomic Physics, Ministry of Research and Innovation and Institute of Atomic Physics and Universitatea Nationala de Stiinta si Tehnologie Politehnica Bucuresti, Romania;
Ministry of Education, Science, Research and Sport of the Slovak Republic, Slovakia;
National Research Foundation of South Africa, South Africa;
Swedish Research Council (VR) and Knut \& Alice Wallenberg Foundation (KAW), Sweden;
European Organization for Nuclear Research, Switzerland;
Suranaree University of Technology (SUT), National Science and Technology Development Agency (NSTDA) and National Science, Research and Innovation Fund (NSRF via PMU-B B05F650021), Thailand;
Turkish Energy, Nuclear and Mineral Research Agency (TENMAK), Turkey;
National Academy of  Sciences of Ukraine, Ukraine;
Science and Technology Facilities Council (STFC), United Kingdom;
National Science Foundation of the United States of America (NSF) and United States Department of Energy, Office of Nuclear Physics (DOE NP), United States of America.
In addition, individual groups or members have received support from:
Czech Science Foundation (grant no. 23-07499S), Czech Republic;
FORTE project, reg.\ no.\ CZ.02.01.01/00/22\_008/0004632, Czech Republic, co-funded by the European Union, Czech Republic;
European Research Council (grant no. 950692), European Union;
ICSC - Centro Nazionale di Ricerca in High Performance Computing, Big Data and Quantum Computing, European Union - NextGenerationEU;
Academy of Finland (Center of Excellence in Quark Matter) (grant nos. 346327, 346328), Finland.

\end{acknowledgement}

\bibliographystyle{utphys}   
\bibliography{bibliography}

\appendix
\newpage

\section{The ALICE Collaboration}
\label{app:collab}
\begin{flushleft} 
\small

S.~Acharya\,\orcidlink{0000-0002-9213-5329}\,$^{\rm 126}$, 
A.~Agarwal$^{\rm 134}$, 
G.~Aglieri Rinella\,\orcidlink{0000-0002-9611-3696}\,$^{\rm 32}$, 
L.~Aglietta\,\orcidlink{0009-0003-0763-6802}\,$^{\rm 24}$, 
M.~Agnello\,\orcidlink{0000-0002-0760-5075}\,$^{\rm 29}$, 
N.~Agrawal\,\orcidlink{0000-0003-0348-9836}\,$^{\rm 25}$, 
Z.~Ahammed\,\orcidlink{0000-0001-5241-7412}\,$^{\rm 134}$, 
S.~Ahmad\,\orcidlink{0000-0003-0497-5705}\,$^{\rm 15}$, 
S.U.~Ahn\,\orcidlink{0000-0001-8847-489X}\,$^{\rm 71}$, 
I.~Ahuja\,\orcidlink{0000-0002-4417-1392}\,$^{\rm 36}$
% ...

\section*{Affiliation Notes}

$^{\rm I}$ Deceased\\
$^{\rm II}$ Also at: Max-Planck-Institut fur Physik, Munich, Germany\\
$^{\rm III}$ Also at: Italian National Agency for New Technologies, Energy and Sustainable Economic Development (ENEA), Bologna, Italy\\
$^{\rm IV}$ Also at: Dipartimento DET del Politecnico di Torino, Turin, Italy\\
$^{\rm V}$ Also at: Yildiz Technical University, Istanbul, T\"{u}rkiye\\
$^{\rm VI}$ Also at: Department of Applied Physics, Aligarh Muslim University, Aligarh, India\\
$^{\rm VII}$ Also at: Institute of Theoretical Physics, University of Wroclaw, Poland\\
$^{\rm VIII}$ Also at: Facultad de Ciencias, Universidad Nacional Autónoma de México, Mexico City, Mexico\\
$^{\rm IX}$ Also at: An institution covered by a cooperation agreement with CERN\\

\section*{Collaboration Institutes}

$^{1}$ A.I. Alikhanyan National Science Laboratory (Yerevan Physics Institute) Foundation, Yerevan, Armenia\\
$^{2}$ AGH University of Krakow, Cracow, Poland\\
$^{3}$ Bogolyubov Institute for Theoretical Physics, National Academy of Sciences of Ukraine, Kiev, Ukraine\\
$^{4}$ Bose Institute, Department of Physics  and Centre for Astroparticle Physics and Space Science (CAPSS), Kolkata, India\\
$^{5}$ California Polytechnic State University, San Luis Obispo, California, United States\\
$^{6}$ Central China Normal University, Wuhan, China\\
$^{7}$ Centro de Aplicaciones Tecnol\'{o}gicas y Desarrollo Nuclear (CEADEN), Havana, Cuba\\
$^{8}$ Centro de Investigaci\'{o}n y de Estudios Avanzados (CINVESTAV), Mexico City and M\'{e}rida, Mexico\\
$^{9}$ Chicago State University, Chicago, Illinois, United States\\
$^{10}$ China Institute of Atomic Energy, Beijing, China\\
$^{11}$ China University of Geosciences, Wuhan, China\\
$^{12}$ Chungbuk National University, Cheongju, Republic of Korea\\
$^{13}$ Comenius University Bratislava, Faculty of Mathematics, Physics and Informatics, Bratislava, Slovak Republic\\
$^{14}$ Creighton University, Omaha, Nebraska, United States\\
$^{15}$ Department of Physics, Aligarh Muslim University, Aligarh, India\\
$^{16}$ Department of Physics, Pusan National University, Pusan, Republic of Korea\\
$^{17}$ Department of Physics, Sejong University, Seoul, Republic of Korea\\
$^{18}$ Department of Physics, University of California, Berkeley, California, United States\\
$^{19}$ Department of Physics, University of Oslo, Oslo, Norway\\
$^{20}$ Department of Physics and Technology, University of Bergen, Bergen, Norway\\
$^{21}$ Dipartimento di Fisica, Universit\`{a} di Pavia, Pavia, Italy\\
$^{22}$ Dipartimento di Fisica dell'Universit\`{a} and Sezione INFN, Cagliari, Italy\\
$^{23}$ Dipartimento di Fisica dell'Universit\`{a} and Sezione INFN, Trieste, Italy\\
$^{24}$ Dipartimento di Fisica dell'Universit\`{a} and Sezione INFN, Turin, Italy\\
$^{25}$ Dipartimento di Fisica e Astronomia dell'Universit\`{a} and Sezione INFN, Bologna, Italy\\
$^{26}$ Dipartimento di Fisica e Astronomia dell'Universit\`{a} and Sezione INFN, Catania, Italy\\
$^{27}$ Dipartimento di Fisica e Astronomia dell'Universit\`{a} and Sezione INFN, Padova, Italy\\
$^{28}$ Dipartimento di Fisica `E.R.~Caianiello' dell'Universit\`{a} and Gruppo Collegato INFN, Salerno, Italy\\
$^{29}$ Dipartimento DISAT del Politecnico and Sezione INFN, Turin, Italy\\
$^{30}$ Dipartimento di Scienze MIFT, Universit\`{a} di Messina, Messina, Italy\\
$^{31}$ Dipartimento Interateneo di Fisica `M.~Merlin' and Sezione INFN, Bari, Italy\\
$^{32}$ European Organization for Nuclear Research (CERN), Geneva, Switzerland\\
$^{33}$ Faculty of Electrical Engineering, Mechanical Engineering and Naval Architecture, University of Split, Split, Croatia\\
$^{34}$ Faculty of Nuclear Sciences and Physical Engineering, Czech Technical University in Prague, Prague, Czech Republic\\
$^{35}$ Faculty of Physics, Sofia University, Sofia, Bulgaria\\
$^{36}$ Faculty of Science, P.J.~\v{S}af\'{a}rik University, Ko\v{s}ice, Slovak Republic\\
$^{37}$ Faculty of Technology, Environmental and Social Sciences, Bergen, Norway\\
$^{38}$ Frankfurt Institute for Advanced Studies, Johann Wolfgang Goethe-Universit\"{a}t Frankfurt, Frankfurt, Germany\\
$^{39}$ Fudan University, Shanghai, China\\
$^{40}$ Gangneung-Wonju National University, Gangneung, Republic of Korea\\
$^{41}$ Gauhati University, Department of Physics, Guwahati, India\\
$^{42}$ Helmholtz-Institut f\"{u}r Strahlen- und Kernphysik, Rheinische Friedrich-Wilhelms-Universit\"{a}t Bonn, Bonn, Germany\\
$^{43}$ Helsinki Institute of Physics (HIP), Helsinki, Finland\\
$^{44}$ High Energy Physics Group,  Universidad Aut\'{o}noma de Puebla, Puebla, Mexico\\
$^{45}$ Horia Hulubei National Institute of Physics and Nuclear Engineering, Bucharest, Romania\\
$^{46}$ HUN-REN Wigner Research Centre for Physics, Budapest, Hungary\\
$^{47}$ Indian Institute of Technology Bombay (IIT), Mumbai, India\\
$^{48}$ Indian Institute of Technology Indore, Indore, India\\
$^{49}$ INFN, Laboratori Nazionali di Frascati, Frascati, Italy\\
$^{50}$ INFN, Sezione di Bari, Bari, Italy\\
$^{51}$ INFN, Sezione di Bologna, Bologna, Italy\\
$^{52}$ INFN, Sezione di Cagliari, Cagliari, Italy\\
$^{53}$ INFN, Sezione di Catania, Catania, Italy\\
$^{54}$ INFN, Sezione di Padova, Padova, Italy\\
$^{55}$ INFN, Sezione di Pavia, Pavia, Italy\\
$^{56}$ INFN, Sezione di Torino, Turin, Italy\\
$^{57}$ INFN, Sezione di Trieste, Trieste, Italy\\
$^{58}$ Inha University, Incheon, Republic of Korea\\
$^{59}$ Institute for Gravitational and Subatomic Physics (GRASP), Utrecht University/Nikhef, Utrecht, Netherlands\\
$^{60}$ Institute of Experimental Physics, Slovak Academy of Sciences, Ko\v{s}ice, Slovak Republic\\
$^{61}$ Institute of Physics, Homi Bhabha National Institute, Bhubaneswar, India\\
$^{62}$ Institute of Physics of the Czech Academy of Sciences, Prague, Czech Republic\\
$^{63}$ Institute of Space Science (ISS), Bucharest, Romania\\
$^{64}$ Institut f\"{u}r Kernphysik, Johann Wolfgang Goethe-Universit\"{a}t Frankfurt, Frankfurt, Germany\\
$^{65}$ Instituto de Ciencias Nucleares, Universidad Nacional Aut\'{o}noma de M\'{e}xico, Mexico City, Mexico\\
$^{66}$ Instituto de F\'{i}sica, Universidade Federal do Rio Grande do Sul (UFRGS), Porto Alegre, Brazil\\
$^{67}$ Instituto de F\'{\i}sica, Universidad Nacional Aut\'{o}noma de M\'{e}xico, Mexico City, Mexico\\
$^{68}$ iThemba LABS, National Research Foundation, Somerset West, South Africa\\
$^{69}$ Jeonbuk National University, Jeonju, Republic of Korea\\
$^{70}$ Johann-Wolfgang-Goethe Universit\"{a}t Frankfurt Institut f\"{u}r Informatik, Fachbereich Informatik und Mathematik, Frankfurt, Germany\\
$^{71}$ Korea Institute of Science and Technology Information, Daejeon, Republic of Korea\\
$^{72}$ Laboratoire de Physique Subatomique et de Cosmologie, Universit\'{e} Grenoble-Alpes, CNRS-IN2P3, Grenoble, France\\
$^{73}$ Lawrence Berkeley National Laboratory, Berkeley, California, United States\\
$^{74}$ Lund University Department of Physics, Division of Particle Physics, Lund, Sweden\\
$^{75}$ Nagasaki Institute of Applied Science, Nagasaki, Japan\\
$^{76}$ Nara Women{'}s University (NWU), Nara, Japan\\
$^{77}$ National and Kapodistrian University of Athens, School of Science, Department of Physics , Athens, Greece\\
$^{78}$ National Centre for Nuclear Research, Warsaw, Poland\\
$^{79}$ National Institute of Science Education and Research, Homi Bhabha National Institute, Jatni, India\\
$^{80}$ National Nuclear Research Center, Baku, Azerbaijan\\
$^{81}$ National Research and Innovation Agency - BRIN, Jakarta, Indonesia\\
$^{82}$ Niels Bohr Institute, University of Copenhagen, Copenhagen, Denmark\\
$^{83}$ Nikhef, National institute for subatomic physics, Amsterdam, Netherlands\\
$^{84}$ Nuclear Physics Group, STFC Daresbury Laboratory, Daresbury, United Kingdom\\
$^{85}$ Nuclear Physics Institute of the Czech Academy of Sciences, Husinec-\v{R}e\v{z}, Czech Republic\\
$^{86}$ Oak Ridge National Laboratory, Oak Ridge, Tennessee, United States\\
$^{87}$ Ohio State University, Columbus, Ohio, United States\\
$^{88}$ Physics department, Faculty of science, University of Zagreb, Zagreb, Croatia\\
$^{89}$ Physics Department, Panjab University, Chandigarh, India\\
$^{90}$ Physics Department, University of Jammu, Jammu, India\\
$^{91}$ Physics Program and International Institute for Sustainability with Knotted Chiral Meta Matter (WPI-SKCM$^{2}$), Hiroshima University, Hiroshima, Japan\\
$^{92}$ Physikalisches Institut, Eberhard-Karls-Universit\"{a}t T\"{u}bingen, T\"{u}bingen, Germany\\
$^{93}$ Physikalisches Institut, Ruprecht-Karls-Universit\"{a}t Heidelberg, Heidelberg, Germany\\
$^{94}$ Physik Department, Technische Universit\"{a}t M\"{u}nchen, Munich, Germany\\
$^{95}$ Politecnico di Bari and Sezione INFN, Bari, Italy\\
$^{96}$ Research Division and ExtreMe Matter Institute EMMI, GSI Helmholtzzentrum f\"ur Schwerionenforschung GmbH, Darmstadt, Germany\\
$^{97}$ Saga University, Saga, Japan\\
$^{98}$ Saha Institute of Nuclear Physics, Homi Bhabha National Institute, Kolkata, India\\
$^{99}$ School of Physics and Astronomy, University of Birmingham, Birmingham, United Kingdom\\
$^{100}$ Secci\'{o}n F\'{\i}sica, Departamento de Ciencias, Pontificia Universidad Cat\'{o}lica del Per\'{u}, Lima, Peru\\
$^{101}$ Stefan Meyer Institut f\"{u}r Subatomare Physik (SMI), Vienna, Austria\\
$^{102}$ SUBATECH, IMT Atlantique, Nantes Universit\'{e}, CNRS-IN2P3, Nantes, France\\
$^{103}$ Sungkyunkwan University, Suwon City, Republic of Korea\\
$^{104}$ Suranaree University of Technology, Nakhon Ratchasima, Thailand\\
$^{105}$ Technical University of Ko\v{s}ice, Ko\v{s}ice, Slovak Republic\\
$^{106}$ The Henryk Niewodniczanski Institute of Nuclear Physics, Polish Academy of Sciences, Cracow, Poland\\
$^{107}$ The University of Texas at Austin, Austin, Texas, United States\\
$^{108}$ Universidad Aut\'{o}noma de Sinaloa, Culiac\'{a}n, Mexico\\
$^{109}$ Universidade de S\~{a}o Paulo (USP), S\~{a}o Paulo, Brazil\\
$^{110}$ Universidade Estadual de Campinas (UNICAMP), Campinas, Brazil\\
$^{111}$ Universidade Federal do ABC, Santo Andre, Brazil\\
$^{112}$ Universitatea Nationala de Stiinta si Tehnologie Politehnica Bucuresti, Bucharest, Romania\\
$^{113}$ University of Cape Town, Cape Town, South Africa\\
$^{114}$ University of Derby, Derby, United Kingdom\\
$^{115}$ University of Houston, Houston, Texas, United States\\
$^{116}$ University of Jyv\"{a}skyl\"{a}, Jyv\"{a}skyl\"{a}, Finland\\
$^{117}$ University of Kansas, Lawrence, Kansas, United States\\
$^{118}$ University of Liverpool, Liverpool, United Kingdom\\
$^{119}$ University of Science and Technology of China, Hefei, China\\
$^{120}$ University of South-Eastern Norway, Kongsberg, Norway\\
$^{121}$ University of Tennessee, Knoxville, Tennessee, United States\\
$^{122}$ University of the Witwatersrand, Johannesburg, South Africa\\
$^{123}$ University of Tokyo, Tokyo, Japan\\
$^{124}$ University of Tsukuba, Tsukuba, Japan\\
$^{125}$ Universit\"{a}t M\"{u}nster, Institut f\"{u}r Kernphysik, M\"{u}nster, Germany\\
$^{126}$ Universit\'{e} Clermont Auvergne, CNRS/IN2P3, LPC, Clermont-Ferrand, France\\
$^{127}$ Universit\'{e} de Lyon, CNRS/IN2P3, Institut de Physique des 2 Infinis de Lyon, Lyon, France\\
$^{128}$ Universit\'{e} de Strasbourg, CNRS, IPHC UMR 7178, F-67000 Strasbourg, France, Strasbourg, France\\
$^{129}$ Universit\'{e} Paris-Saclay, Centre d'Etudes de Saclay (CEA), IRFU, D\'{e}partment de Physique Nucl\'{e}aire (DPhN), Saclay, France\\
$^{130}$ Universit\'{e}  Paris-Saclay, CNRS/IN2P3, IJCLab, Orsay, France\\
$^{131}$ Universit\`{a} degli Studi di Foggia, Foggia, Italy\\
$^{132}$ Universit\`{a} del Piemonte Orientale, Vercelli, Italy\\
$^{133}$ Universit\`{a} di Brescia, Brescia, Italy\\
$^{134}$ Variable Energy Cyclotron Centre, Homi Bhabha National Institute, Kolkata, India\\
$^{135}$ Warsaw University of Technology, Warsaw, Poland\\
$^{136}$ Wayne State University, Detroit, Michigan, United States\\
$^{137}$ Yale University, New Haven, Connecticut, United States\\
$^{138}$ Yildiz Technical University, Istanbul, Turkey\\
$^{139}$ Yonsei University, Seoul, Republic of Korea\\
$^{140}$ Affiliated with an institute covered by a cooperation agreement with CERN\\
$^{141}$ Affiliated with an international laboratory covered by a cooperation agreement with CERN.\\

\end{flushleft} 

\end{document}